\documentclass[12pt,a4paper]{article}
\usepackage{arxiv}

\usepackage[utf8]{inputenc}
\usepackage[T1]{fontenc}    
\usepackage{hyperref}       
\usepackage{url}            
\usepackage{booktabs}       
\usepackage{amsfonts}       
\usepackage{amsmath}
\usepackage{amssymb}
\usepackage{nicefrac}       
\usepackage{microtype}      
\usepackage{cleveref}       
\usepackage{lipsum}         
\usepackage{graphicx}
\usepackage{subfig}
\usepackage{natbib}
\usepackage{doi}
\usepackage{xcolor}
\usepackage{bm}

\title{Implications of Modeling Seasonal Differences in the Extremal Dependence of Rainfall Maxima}

\date{}

\author{ \href{https://orcid.org/0000-0001-6727-4776}{\includegraphics[scale=0.06]{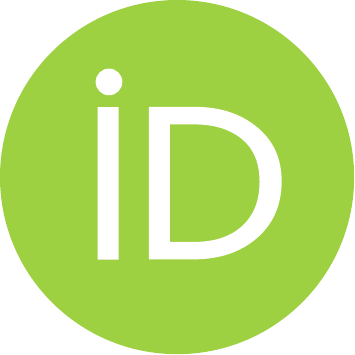}\hspace{1mm}Oscar E.~Jurado}\thanks{Corresponding author} \\
	Institut für Meteorologie\\
	Freie Universität Berlin\\
	12165 Berlin, Germany \\
	\texttt{jurado@zedat.fu-berlin.de} \\
	\And
	\href{https://orcid.org/0000-0001-5978-3204}{\includegraphics[scale=0.06]{orcid.pdf}\hspace{1mm}Marco ~Oesting} \\
	Stuttgart Center for Simulation Science (SC SimTech) \& Institute for Stochastics and Applications\\
	University of Stuttgart\\
	70569 Stuttgart, Germany\\
	\texttt{marco.oesting@mathematik.uni-stuttgart.de} \\
	\And
	\href{https://orcid.org/0000-0003-0763-3954}{\includegraphics[scale=0.06]{orcid.pdf}\hspace{1mm}Henning W.~Rust} \\
	Institut für Meteorologie\\
	Freie Universität Berlin\\
	12165 Berlin, Germany\\
	\texttt{henning.rust@fu-berlin.de} \\
}

\renewcommand{\shorttitle}{Seasonal Spatial Dependence}

\hypersetup{
pdftitle={Implications of Modeling Seasonal Differences in the Extremal Dependence of Rainfall Maxima},
pdfauthor={Oscar E.~Jurado, Marco ~Oesting, Henning W.~Rust},
pdfkeywords={max-stable process, extreme rainfall modeling, Bayesian statistics},
}

\begin{document}

\maketitle

\begin{abstract}
For modeling extreme rainfall, the widely used Brown-Resnick
  max-stable model extends the concept of the variogram to suit block
  maxima, allowing the explicit modeling of the extremal dependence
  shown by the spatial data. This extremal dependence stems from the
  geometrical characteristics of the observed rainfall, which is
  associated with different meteorological processes and is usually
  considered to be constant when designing the model for a
  study. However, depending on the region, this dependence can change
  throughout the year, as the prevailing meteorological conditions
  that drive the rainfall generation process change with the
  season. Therefore, this study analyzes the impact of the seasonal
  change in extremal dependence for the modeling of annual block
  maxima in the Berlin-Brandenburg region. For this study, two seasons
  were considered as proxies for different dominant meteorological
  conditions: summer for convective rainfall and winter for
  frontal/stratiform rainfall. Using maxima from both seasons, we
  compared the skill of a linear model with spatial covariates (that
  assumed spatial independence) with the skill of a  Brown-Resnick
  max-stable model. This comparison showed a considerable difference
  between seasons, with the isotropic Brown-Resnick model showing
  considerable loss of skill for the winter maxima. We conclude that
  the assumptions commonly made when using the Brown-Resnick model are
  appropriate for modeling summer (i.e., convective) events, but
  further work should be done for modeling other types of
  precipitation regimes.
\end{abstract}



\section{Introduction}\label{sec1}

The statistical modeling of extreme precipitation is essential for
designing public hydrological infrastructure and urban planning
worldwide \citep{durrans}. This approach typically combines observed
information from past events with models from Extreme Value Theory
(EVT) to give a probabilistic estimate of the magnitude and frequency
of future extreme precipitation events \citep{Coles2001}. Information
about past events usually comes from ground observations (e.g., rain
gauges), operated mainly by local weather services. For a typical EVT
application, information from rain gauges is used to fit the
parameters of a max-stable distribution (such as the Generalized
Extreme Value (GEV) distribution), from which information on the
magnitude and frequency of events in the far-right tail of the
distribution can be elicited. The ultimate goal of EVT analyses is
then to provide adequate estimates of these estimates along with their
uncertainties.  These estimates are commonly communicated to
decision-makers either in the form of return periods for certain
return levels (i.e., ``1-in-n years event'') or as a more general
quantity like the probability of exceedance and risk of failure over a
given design life period \citep{Serinaldi2015,Rootzen2013}. 

A common problem when modeling extreme rainfall is that no
observations exist in many locations where information from
statistical modelling of extreme events would be useful.  However, on
many occasions, observations exist near unobserved locations.  This
setting is the same as in Geostatistics, except that the focus is on
extremes and max-stable distributions in this case.  This problem has
given way to different EVT models that allow interpolation of
estimates to unobserved locations, usually englobed within the term
``Spatial Extremes''. Spatial Extremes models follow a very similar
theoretical background to the methods of Geostatistics and can be
thought of as extensions of Geostatistics, but for extremes
\citep{Davison2012a}.

Most Spatial Extremes and Geostatistical models use the so-called
first law of Geography: ``everything is related to everything else,
but near things are more related than distant things.''
\citep{tobler1970}.  In other words, there exists a particular
covariance function that depends on the distance between points with
observations.  Spatial models use the observations from the different
locations to fit a covariance function that describes how much two or
more variables change as a function of some distance metric.  Thus,
covariance functions describe the spatial dependence between the
observed locations. In the case of Spatial Extremes, the corresponding
analog to the covariance function (e.g., the tail-dependence function)
is combined with an appropriate model for extremes to fit a joint
distribution for the different locations and, in some cases, to also
obtain the estimates of the marginal parameters in each
location. Interpolation to unobserved locations is then achieved by
combining the fitted tail-dependence function with the fitted model.

When dealing with block maxima stemming from observations fixed in
space (e.g., rain gauges), a commonly used spatial extremes model is a
max-stable process \citep{Davison2015}. Max-stable processes are an
extension to infinite dimensions of univariate EVT models for block
maxima \citep{Padoan2010}. Unlike univariate EVT models, there does
not exist a single parametric family of max-stable processes to which block
maxima always converge. Nevertheless, diverse parametric families with
different tail-dependence functions have been proposed. For the
spatial modeling of extreme precipitation, a commonly used family of
max-stable processes is the Brown-Resnick family
\citep{Le2018,Davison2012,Buhl2016}. Brown-Resnick models are based on
Gaussian processes with a tail dependence function that includes the
geostatistical concept of the (semi-)variogram. Assuming that the
underlying Gaussian process possesses stationary increments
(i.e., it is only a function of the distance between different
stations), the spatial dependence structure can be modeled exclusively
with the variogram.

In previous studies using Brown-Resnick models for extreme rainfall,
the focus has been on maxima that stem from summer events. This choice
is typically justified as rainfall events in summer are usually the
events with the largest magnitude and, thus, the ones with the most
significant impact.  Furthermore, these events are usually associated
with convective activity, which for the study region is predominant in
summer \citep{Berg2013}. Nevertheless, little work has been done to
model extreme rainfall resulting from other types of events, such as
stratiform ones. These events are relevant, as they could be the
dominant types in other regions of the planet or of interest to
different stakeholders. An essential aspect of our study is that these
events differ significantly in terms of spatial and temporal extent,
which likely leads to different spatial dependence structures,
creating a need to research and improve our understanding of modeling
extreme rainfall for maxima that originates from different types of
events.

The present study aims to investigate how the extremal dependence
changes for different rainfall-generating mechanisms and how this
change influences the estimation of return levels down the line.  We
use a Brown-Resnick max-stable process to model the spatial
variability of precipitation maxima in the Berlin-Brandenburg region,
accounting for the extremal dependence. We apply this model to
half-year block maxima from two seasons, winter and summer. We
hypothesize that summer block maxima come mainly from convective events, while winter block
maxima come from stratiform and frontal events. We then approximate
how the dependence changes with different rainfall-generating
mechanisms using these two kinds of semi-annual block maxima.
Furthermore, we selected two temporal scales for each season to
investigate the impact of processes with different time scales. We fit
a Bayesian distributional linear model that assumes
independence in space as a reference to our spatial model to discern
the effects of the change in dependence on the estimated return
levels.  This reference model is compared with the spatial model
within a verification framework.

This study is organized as follows: first, we present a review of the
different types of rainfall-generating mechanisms that dominate in our
study region. Then, we present the EVT methods we used to model
extreme rainfall.  Afterward, we introduce a verification framework to
compare the different models, from which we present the results to
determine whether a considerable change in the extremal dependence was
observed and their consequences on the reported return levels.


Rainfall is the result of complex processes and interactions in the
hydro- and atmosphere, involving processes from a wide range of
temporal and spatial scales. In particular, for the midlatitude
region, rainfall characteristics are heavily associated with the
synoptic weather situation present when the event
happened. \cite{Walther2006} classifies rainfall events as either
frontal or convective based on synoptic-scale considerations. The
synoptic-scale is usually defined as a length scale of around 2000 km
to 20000 km, involving events that last from days to weeks. The
distinction between synoptic or convective
rainfall is relevant for the statistical modeling efforts, as rainfall
events associated with fronts (i.e., synoptic-scale) have very
different temporal and spatial characteristics than those associated
with isolated convective (typically mesoscale) events. By way of
illustration, \cite{Orlanski1975} characterizes thunderstorms as
events lasting from half an hour up to a few hours covering areas of
several km$^2$, and frontal events as having a lifetime of more than a
day with a spatial spread of hundreds of km. These spatiotemporal
characteristics can also influence the magnitude and the timing of
extreme rainfall events; for example, \cite{Bohnenstengel2011} found
that for a 25 km x 25 km region to the southeast of Berlin, extreme
precipitation events occur more often in times of convective events
than during times with frontal precipitation.

\begin{figure}
\centering
\includegraphics[width=0.8\textwidth]{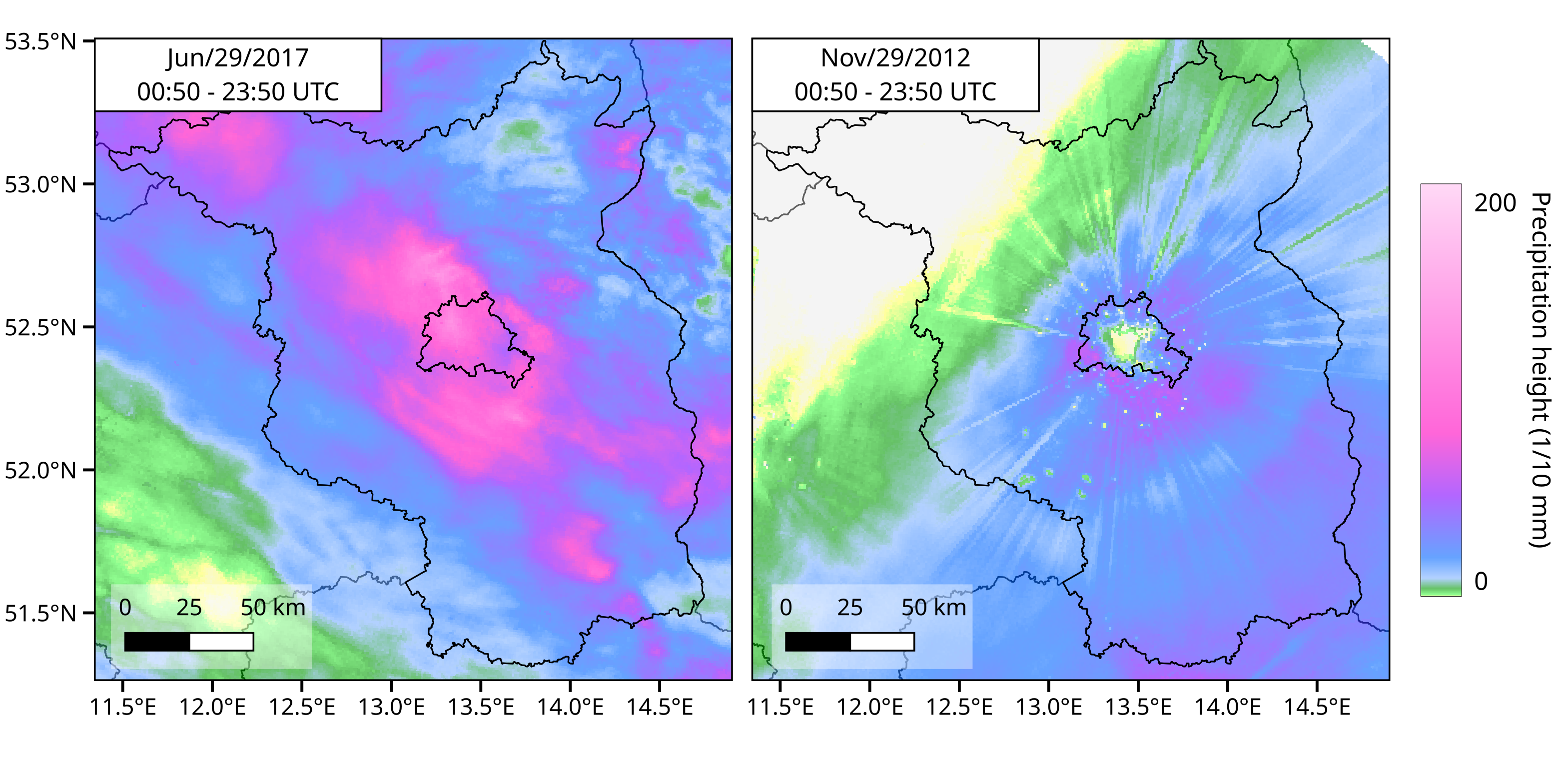}
\caption{Map showing daily accumulated precipitation height for two 
extreme precipitation events chosen arbitrarily for demonstration). 
Left: A summer convective event. Right: A winter frontal event. Data comes 
from the RADOLAN database made available from the DWD.}
\label{fig:radar_map}
\end{figure}

In the midlatitudes region, the type of dominant rainfall-generating
mechanism changes during the year. For example, \cite{Berg2013} found
that synoptic observations of convective events dominated during the
summer seasons in four stations across Germany. In contrast, they
found that most rainfall in the winter months resulted from stratiform
clouds (commonly associated with frontal events).  Thus, we predict
that when looking at seasonal block maxima for a study region in
Germany, summer maxima will originate mainly from convective events,
while winter maxima will primarily originate from frontal ones.  An
example of this can be seen in Fig.~\ref{fig:radar_map}, which shows
the daily precipitation height in the Berlin-Brandenburg region for a
convective event in summer (left) against that of a frontal/stratiform
event in winter (right). For most stations in the domain, the
semi-annual block maxima in the two corresponding seasons were attained
for these two particular events, 
meaning they can be seen as extreme events.  Extremal dependence in
space arises when an extreme event is large enough to impact several
rain gauges simultaneously. Therefore, the extremal dependence heavily
depends on the spatial characteristics presented by the rainfall
generating mechanisms. Thus, if these mechanisms change seasonally
during the year, we expect the dependence structure to also change
throughout the year.


\section{Methods and Data}

In this study, we perform the statistical modeling of extreme rainfall
using the block maxima approach. This approach is based on the
Fisher-Tippett-Gnedenko theorem, which states that under mild conditions
block maxima of a sufficiently large block length of independently and
identically distributed random variables can be approximately
modeled by the generalized extreme value (GEV) distribution. When
dealing with rainfall, block lengths of one month have been proven to
be long enough to
assure convergence to the GEV distribution \citep{Fischer2017}.  Yearly block maxima can
then be used to fit the parameters of the GEV for each rain gauge
individually, resulting in the ``zeroth-order'' approach to modeling
extreme rainfall in space.  This pointwise approach, however, does not
pool any information across stations and, therefore, cannot predict
values for ungauged sites. Prediction of ungauged sites can be
achieved by extending the pointwise GEV approach to include spatial
covariates, which pools information from different locations,
typically resulting in reduced uncertainties for the estimated
parameters of the GEV \citep{Ulrich2020}.  Nevertheless, this second
approach ignores the spatial dependence in the data, resulting in a
misspecified likelihood that consistently underestimates the
uncertainty of the estimates.  Our study extends this approach by
using a max-stable process to include spatial dependence.

\subsection{Data}

\begin{figure}
\centering
\includegraphics[width=0.5\textwidth]{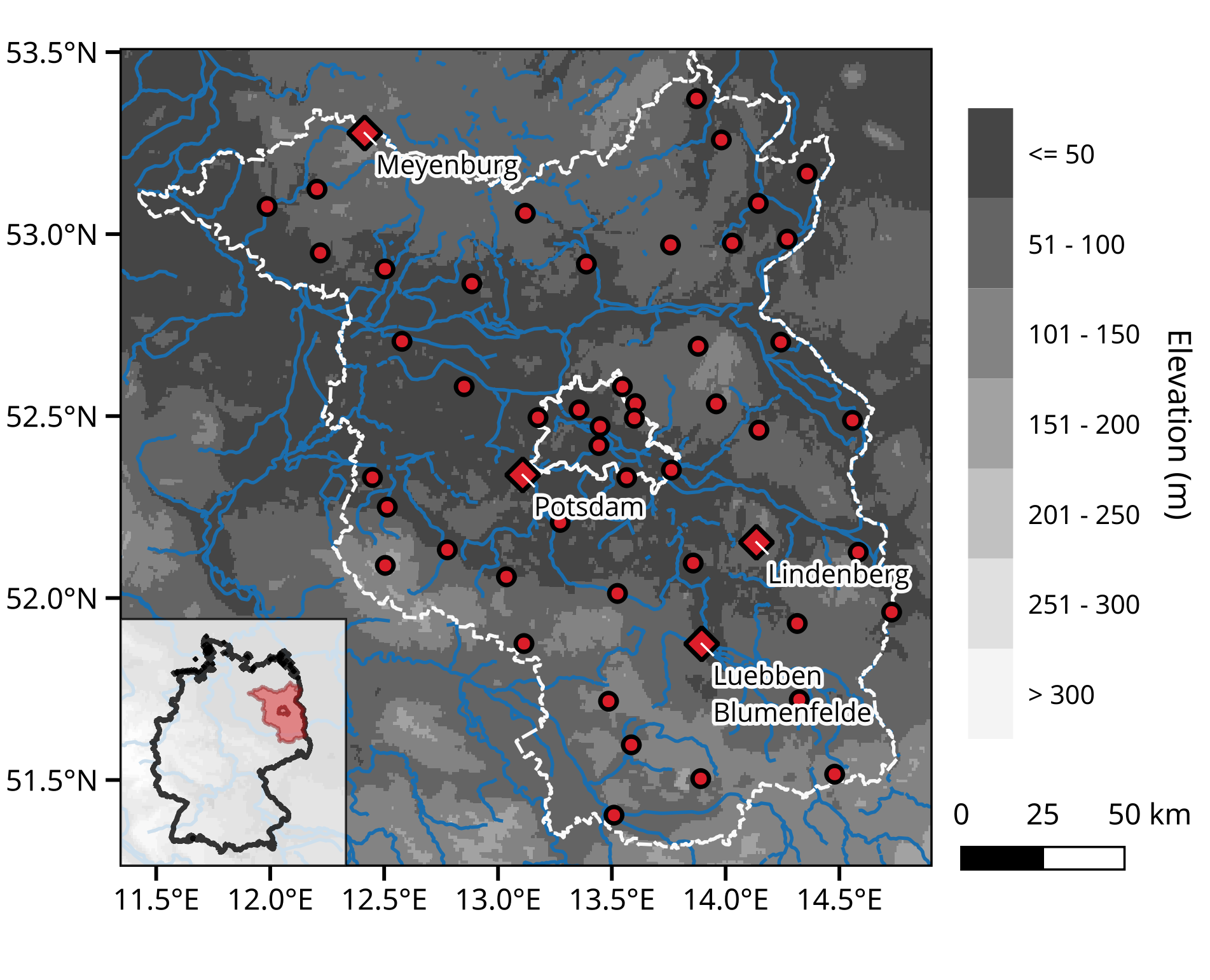}
\caption{Map showing the location
inside the Berlin-Brandenburg area of the DWD weather stations included
in this study (red dots). The lower left inset shows the location of the study
domain within Germany. Diamonds show the reference stations used to
showcase results. }
\label{fig:BB_map}
\end{figure}

We used accumulated hourly and daily precipitation height measurements
(in mm) from 53 stations belonging to the German Meteorological
Service (DWD) in the Berlin-Brandenburg region of Germany
(Fig.~\ref{fig:BB_map}). The data was acquired through the German
Meteorological Service (DWD) Open Data Server using the
\texttt{R}-package \texttt{rdwd} \citep{rdwd}. The stations were chosen
to include only those that contained measurements with both hourly and
daily periods. This choice reduced the available number of stations
with daily measurements from 300 to 53. Reducing the total number of
stations was considered necessary to ensure the fairness of the
comparisons with results using stations with hourly measurements and
to lower the computational burden needed to fit the models. The
average distance between all station pairs was approx.\ 95 km, with a
range of \([4,245]\) km. The raw data contains further information
about the type of precipitation measured (liquid or solid), but for
the purpose of this study no discrimination was done with regard to
precipitation type.

Two different periods were considered for this study: from 1970-2020
for the daily observations and 2004-2020 for the hourly
observations. These periods were chosen in order to minimize the
number of invalid pairs when using the pairwise likelihood (see Appendix~\ref{appendix:a}).

At the location $s_j \in \mathcal{S}$, where $\mathcal{S}$ represents the geographical domain and $j=1,...,n$ is an index denoting the rain gauge, the data contains the accumulated rainfall values $(r_{d,1}(s_j),...,r_{d,k_j}(s_j))$ in mm, where $d \in \{1,24\}$ is an index for the duration of the considered precipitation events, namely, hourly or daily. Different gauges can have different lengths for the measurement period, so that $k_j$ depends on the location $s_j$. The accumulated rainfall values were transformed to the average hourly/daily intensity values $(\zeta_{d,1}(s_j),...,\zeta_{d,k_i}(s_j))$ in mm/h.

Following \citep{Koutsoyiannis1998}, the average hourly intensity data $\zeta_{d=1,\tau}(s)$ (where $\tau$ represents the time (in hours) of the observation) were aggregated
to create the 12-hour accumulated precipitation intensity time series $\zeta_{d=12,\tau}(s)$ (in mm/h). This aggregation was necessary because a visual inspection
of the pairwise extremal coefficient resulting from the hourly series
strongly suggested that the data was asymptotically independent, which
violates a major assumption for using max-stable processes. The lowest
aggregation duration that did not show asymptotic independence was
12 hours. The 12-hour aggregated series is obtained using
\begin{equation}
\zeta_{d=12,\tau}(s) = \frac{1}{12} \sum_{i=0}^{11} \zeta_{d=1,\tau-i} (s),
\label{eq:agg}
\end{equation}
which can be seen as a moving average with a time window of 12
hours. The aggregation described in Eq. \eqref{eq:agg} was done using
the package \texttt{IDF} \citep{idf}.


The 12-hour $\bm{\zeta}_{12} (s_j)$ and daily $\bm{\zeta}_{24} (s_j)$ average
precipitation intensity series are then used to get four series of
semi-annual block maxima series $(i_{d,t=1}^{l}(s_j),...,i_{d,t=N_j}^{l}(s_j))$. In this case, the index $t$ can be seen as indicating the year.  These four
series result from combining the two durations $d \in \{12,24\}$
and the two seasons $l \in \{\mathrm{sum},\mathrm{win}\}$  using the corresponding abbreviations for sumer and winter, respectively. The semi-annual block maxima were
obtained using 
\begin{equation}
i_{d,t}^l(s) = \max_{l_t^- < \tau < l_t^+} {\zeta_{d,\tau} (s)},
\end{equation}
where \(l_t^-\) and \(l_t^+\) correspond to the beginning and end of
either winter or summer for each year $t$. For this work, we
consider summer as May, June, July, and August; winter is considered
to be the months of January, February, November, and December. In
order to avoid having winter block maxima that come from disconnected
months, we shifted the $\bm{\zeta}_d (s)$ values of November and
December to the following year, making the four winter months of any
given calendar year come from the same ``meteorological'' winter.
Note that for each instance of $\tau$ within the same type of season,
i.e.\ summer or winter, and for each fixed duration $d \in \{12,24\}$, 
we perceive $\zeta_{d,\tau}(\cdot)$ as independent realizations of some
stochastic process $\{X(s): \ s \in \mathcal{S}\}$ which will be the justification
for the use of GEV distributions and max-stable processes for modeling
the distribution of $i_{d,t}^l(s)$ below.

\begin{figure}
\centering
\includegraphics[width=\textwidth]{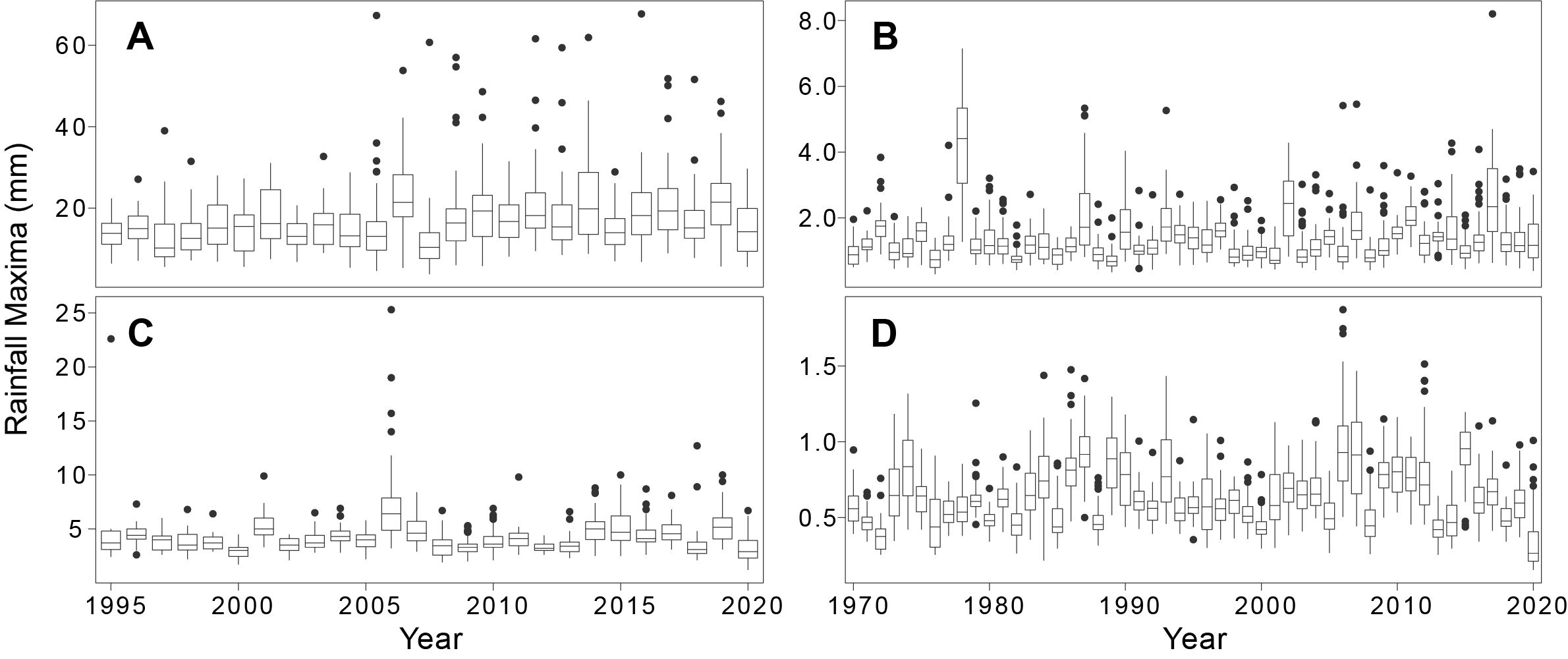}
\caption{Boxplots showing the distribution of the rainfall semi-annual
block maxima for the 53 stations included in this study. 
(A) 12-hour summer maxima, (B) daily summer maxima,
(C) 12-hour winter maxima, and (D) daily winter maxima.}
\label{fig:boxplots_bmax}
\end{figure}

Figure~\ref{fig:boxplots_bmax} shows the temporal distribution of the
semmi-annual block maxima for
summer, i.e.\ $\bm{i}^{\mathrm{sum}}_{12}(s)$ and $\bm{i}^{\mathrm{sum}}_{24}(s)$, 
and winter, i.e.\ $(\bm{i}^{\mathrm{win}}_{12}(s)$ and $\bm{i}^{\mathrm{win}}_{24}(s))$ over the 53 stations. From
Figure~\ref{fig:boxplots_bmax}, it is apparent that the magnitude of
the maxima changes depending on the season, with consistently larger
values for summer events. The final length of the daily series is 50
years, while for the 12-hour series, it is 26 years.

\subsection{Characterizing extremal dependence}

To explore how the bivariate extremal dependence changes for the block
maxima derived from the summer and winter seasons, we used an estimate
of the empirical pairwise extremal coefficient $\theta(s_j,s_{j'})$, which is a
summary measure of dependence of a random
two-dimensional vector $(X(s_j),X(s_{j'}))$ \citep{Coles2001,dey2016_spext}. The pairwise
extremal coefficient can take values in the rage \([1,2]\), where 1
denotes complete dependence and 2 asymptotic independence.

For each pair $(s_j,s_{j'})$ of locations of gauged stations, we estimate the empirical extremal coefficient $\hat\theta_{\text{NP}}(s_j,s_{j'})$
using the non-parametric method proposed by \cite{Marcon2017}, which
\citep{Vettori2018} found to have the best overall performance
compared to other empirical estimators. The estimation of
$\hat\theta_{\text{NP}}(s_j,s_{j'})$ is done with the \texttt{R}-package
\texttt{ExtremalDep} \citep{extremaldep}. This method requires
the specification of a polynomial order, for which a graphical
analysis (not shown) found that
a fixed value of \(k=20\) yielded the most appropriate values of
$\hat\theta_{\text{NP}}(s_j,s_{j'})$ for the
different \(\bm{i}^l_{d}(s)\) series.

\subsection{Modeling of extreme rainfall}
\label{sec:modelling-extreme-rainfall}

We follow a two-step approach to model the $\bm{i}^l_{d}(s)$
series. In the first step, we model the marginal distribution of the
pooled data from all stations by including spatial covariates within a
Bayesian distributional model (DM). 
For the second step, we extend the model of the first step
with a max-stable process, allowing the model to capture the so-called
``residual dependence'' left from the first-step that arises
from the extremal dependence \citep{Cooley2012}. We then compare the
models from both steps using a forecast verification framework to
study how the extremal dependence influences the estimates of the
model parameters. We consider the BDM approach to act as a ``control''
compared to the max-stable process approach, allowing us to explore
how the seasonal difference in the extremal dependence affects the
estimates down the line.

Estimations made within the framework of extreme value statistics are
usually made with small data samples, as extreme events are by
definition rare. The small sample size, in turn, leads to high
uncertainty of all estimates, a problem compounded by the fact that
most applications of EVT focus on the very far right of the
distribution, where estimates already have high levels of
uncertainty. Therefore, any EVT study must include information about
the uncertainty that can be easily interpreted and adapted for the
final-user applications.  Uncertainty in this study is exclusively
obtained using Bayesian methods for inference, which allow a
straightforward and intuitive interpretation of their values.

The following sections explore the two approaches used for this study:
First, the approach that includes spatial covariates but assumes
independence in space (henceforth denoted as the DM approach),
and second, the approach that uses a Brown-Resnick max-stable process
to account for the spatial dependence (henceforth denoted as the BR
approach).

\subsubsection{Using a Bayesian distributional model}\label{sec:independence}

A simple but effective approach to model the variability of extreme
rainfall in space is to pool information from all stations in the
study region and assume that all values are independent and
identically distributed. This approach assumes that
observations at each station are independent of those at any
other station. 
Instead, information is pooled from different stations using spatial
covariates, such as the position of each station, as a predictor
within a model. The resulting model can then 
characterize extremal behavior at unobserved
locations simply by using their position in the covariates.

In this study, we use an analog of Vector Generalized Linear Models
known in the Bayesian literature as Distributional Models (DMs), or sometimes,
as Bayesian distributional regression \citep{Umlauf2018}.  
Distributional models allow for the simultaneous linear modeling of all
distributional parameters. This is in contrast to standard GLMs, where only the
location parameter is modeled. Furthermore, like VLGMs, DMs allow the use 
of distributions from outside the exponential family, such as the GEV distribution.
Extending a GLM to be a Bayesian DM is straightforward, as one requires
only to add the additional log-likelihood contribution from the
additional parameter models in the MCMC steps. Using these models, we
can incorporate spatial covariates into linear models for every
parameter of the marginal distributions.

For every rain gauge $j$ located at $s_j$, the block maxima $\bm{i}^l_{d}(s_j) = (i^l_{d,1}(s_j),...,i^l_{d,N_j}(s_j))$ are
assumed to be i.i.d.\ and, as the Fisher-Tippett-Gnedenko Theorem for block maxima suggests, follow the Generalized Extreme Value (GEV)
distribution, which following \citep{Coles2001} is given by 
\begin{equation}
G(x) = \begin{cases} \exp \left[-\left(1+\xi \frac{x - \mu}{\sigma} \right)_+^{-1/\xi} \right] &
 \xi \neq 0, \\ 
 \exp\left[ - \frac{x-\mu}{\sigma}\right] &
 \xi = 0, \end{cases}
\label{eq:gev}
\end{equation}
where $\mu \in \mathbb{R},\sigma>0,\xi \in \mathbb{R}$ are the location, scale, and shape
parameters, respectively, and \(x_{+} = \max(0,x)\). This assumption
is verified for all stations using Quantile-Quantile plots (not shown).

We then follow \citep{Fischer2017} and describe the spatial
  variation of location $\mu$ and scale $\sigma$ using a linear
  combination of Legendre polynomials of longitude and latitude as
  covariates. Legendre polynomials form a set of orthogonal basis
functions on \([-1,1]\), ensuring that their evaluations at the covariates -- normalized to that interval --  
will be linearly independent. Our model is restricted only to the northing and easting
coordinates, ignoring the altitude. Thus, we are left with the
distributional model
\begin{align}
\mu(s) &= \beta_0^\mu + \sum_{j=1}^J \beta_{j,x}^\mu P_j(x') + \sum_{k=1}^K \beta_{k,y}^\mu P_k(y'), \label{eq:glm_mu}\\
\log(\sigma(s)) &= \beta_{0}^\sigma + \sum_{j=1}^J \beta_{j,x}^\sigma P_j(x') + \sum_{k=1}^K \beta_{k,y}^\sigma P_k(y'), \label{eq:glm_sigma}\\
\xi &= \xi\, , \label{eq:glm_xi}
\end{align}
where $s=(x',y')$, and a logarithmic link function is used for the scale parameter $\sigma$ to ensure
positivity. \(P_i(\cdot)\) denotes the \(i\)th order Legendre
Polynomial. We transform the coordinates from longitude and
latitude to \emph{Universal Transverse Mercator} (UTM) \(x\) and \(y\)
coordinates (UTM zone 33N) so that the distances between stations are
measured in meters instead of degrees, simplifying the analysis. The
$(x,y)$ coordinates are then shifted and scaled to the \((x',y')\)
coordinates within the \([-1,1] \times [-1,1]\) square in order to
compute the respective Legendre Polynomials.

The shape parameter \(\xi\) is left constant throughout the domain, as
other studies have found that this parameter is complicated to
estimate properly and can strongly impact the model's performance
\citep{Cooley2012}.

\paragraph{Model Selection}\label{sec:model-selection}

The linear model in Eqs.~\eqref{eq:glm_mu} and \eqref{eq:glm_sigma}
requires an order for the Legendre Polynomials to be specified. The
order is chosen within the model selection framework using the Widely
Applicable Information Criteria (WAIC) \citep{Vehtari2017}.  A total
of 140 possible combinations of up to order $P_5(\cdot)$ were fitted, and the
model with the lowest WAIC value was chosen.  Furthermore, a
regularizing prior (detailed below) was used to lower the risk of
overfitting.

\subsubsection{Using a max-stable process}\label{sec:dependence}

For the second step of our study, we expanded the model for the
  marginal distribution presented in section \ref{sec:independence}
by a simple max-stable process. The latter was chosen to capture
the extremal dependence in the rainfall maxima. Max-stable processes
are extensions to infinite dimensions of finite-dimensional Extreme
Value Theory models, arising as ``the pointwise maxima taken over an
infinite number of (appropriately rescaled) stochastic processes''
\citep{Ribatet2013}.

More precisely, let $X(s)$ be a random variable representing the daily precipitation
height at site $s \in \mathcal{S}$ (for some fixed duration $d$ and season $l$); that is
$\{X(s): \ s \in \mathcal{S}\}$ is a stochastic process modeling the precipitation at each
site in the spatial domain $\mathcal{S}$. If we have i.i.d.\ replicates $\{X_i(s): \ s \in \mathcal{S}\}$
of the process such as precipitation heights for different days within the same season,
then as already discussed, under mild conditions, the Fisher-Tippett-Gnedenko Theorem states 
that, for each site $s$ and sufficiently large $n$, the distribution of $ \max_{i=1,...,n} X_i(s)$ 
may be approximated by a GEV distribution with spatially varying parameters
$\mu(s), \sigma(s), \xi(s)$. Assuming that not only the marginal distributions, but also the spatial
dependence structures converge, by a spatial extension of the Fisher-Tippett-Gnedenko theorem the block maxima
process $\{ \max_{i=1,...,n} X_i(s): \ s \in \mathcal{S}\}$ can be approximated by a max-stable process
$\{Z'(s): \ s \in \mathcal{S}\}$, given that $n$ is large enough \citep{dey2016_spext}.  
Thus, max-stable processes does not only allow for arbitrary GEV marginal distributions 
$Z'(s) \sim \mathrm{GEV}(\mu(s), \sigma(s), \xi(s))$, but also provide a flexible way of modeling the
dependence structure of the maxima of the $X_i$ random fields.

Consequently, we will assume that the semi-annual block maxima $\bm{i}^l_{d}(s_j)$ form realizations of a max-stable process $\{Z'(s): \ s \in \mathcal{S}\}$ at the gauged sites $s_j \in \mathcal{S}$. It is common in extreme value theory to transform the original block maxima data $\bm{i}^l_{d}(s_j)$ into standardized maxima $\bm{z}^l_{d}(s_j)$ following unit Fréchet marginal distributions (i.e., the case where $\mu = \sigma = \xi = 1$ in Eq.~\ref{eq:gev}). Transformation of the margins to the unit Fréchet distribution does not affect the dependence structure. This transformation is performed via the
relationship
\begin{equation}
z^l_{d,t}(s) = \left[1 + \xi \left(i^l_{d,t}(s) - \frac{\mu^l(s)}{\sigma^l(s)}\right)\right]_+^{1/\xi}.
\end{equation}
As this transformation can be easily reversed, it then allows us to focus on the max-stable process
$\{Z(s): \ s \in \mathcal{S}\}$ without any loss of generality. For such standardized max-stable processes, a variety of parametric submodels has been developed including the popular Brown-Resnick max-stable process model \citep{Kabluchko2009}.

In our study, the marginal standardization requires the specification of response surfaces for
$\mu^l(s)$ and $\sigma^l(s)$ to link $\bm{z}^l_{d}(s)$ to
$\bm{i}^l_{d}(s)$. We chose the response surfaces to have the same
expressions as the model resulting from the model selection of
Eqs.~\eqref{eq:glm_mu} and \eqref{eq:glm_sigma}, with the shape
parameter assumed to be constant over the entire domain.

In theory, max-stable process models can be used to model the joint distribution 
of all semi-annual block maxima $(i^l_{d,1}(s_j),\ldots,i^l_{d,N_j}(s_j))$, $j=1,\ldots,53$, and their standardized analogues $(z^l_{d,1}(s_j),\ldots,z^l_{d,N_j}(s_j))$, $j=1,\ldots,53$, respectively. In practice, however, the resulting likelihood
terms are intractable for even relatively low-dimensional settings. This is why a common
strategy is to restrict the process to the bivariate case, where the distribution functions
and their corresponding densities are well-known. The bivariate
joint probability for the rescaled maxima
\(Pr\{Z^l_{d}(s) \leq z_1, Z^l_{d}(s + \bm{h}) \leq z_2\}\) is then modeled
using the bivariate distribution of the Brown-Resnick max-stable
process model (\cite{Kabluchko2009}, see
appendix~\ref{appendix:a}). For the Brown-Resnick model, the extremal
spatial dependence is a function only of the variogram $\gamma$,
which, with a slight abuse of notation, for this study has the following theoretical model: 
\begin{equation}
\gamma(s, s + \bm{h}) = \gamma(\bm{h}) = \left(\frac{\|\bm{h}\|}{\rho}\right)^\alpha,
\label{eq:variogram}
\end{equation}
Here, $\|\bm{h}\|$ is
the Euclidean distance between the two locations considered,
\(\rho\) is the range parameter, and \(\alpha\) is the smoothness
parameter. The range parameter \(\rho\) can be seen as the distance
for which the dependence is still effective and takes values
\((\rho > 0)\). The smooth parameter \(\alpha\) has no straightforward
interpretation and is constrained to be \(\alpha \in [0,2]\). For this
study, we restrict the variogram to be isotropic and
stationary, i.e.\ $\gamma(s, s + \bm{h})$ depends on $\|\bm{h}\|$ only. The two Brown-Resnick parameters \((\alpha,\rho)\) contain
the information regarding the pairwise dependence structure.
To compare the Brown-Resnick dependence estimated from the model
with the empirical dependence shown by the data, we also obtain a parametric estimate of 
the bivariate extremal coefficient $\theta(s_j,s_{j'})$ using the $(\rho,\alpha)$ parameters
of the Brown-Resnick model. This parametric estimate
$\theta_{\mathrm{BR}}(s_j,s_{j'})$ is computed from the Brown-Resnick variogram
\(\gamma(h)\) obtained in Eq.~\eqref{eq:variogram} using the following relationship:
\begin{equation}
\theta_{\mathrm{BR}}(s_j,s_{j'}) = 2 \Phi (\gamma(\|s_j-s_{j'}\|) / 2)^{1/2}),
\end{equation}
where \(\Phi\) represents the standard normal distribution function.

\subsubsection{Statistical Inference}\label{sec:inference-for-both-models}

The estimation of the posterior distribution of the
$(\beta_0, \beta_{i,P})$ coefficients for the DM approach and the response
surfaces, as well as for the BR dependence parameters
\((\alpha,\rho)\), was carried out using Bayesian inference.  Given a
random variable or vector $Y$ and a probabilistic distribution function
$G(\phi)$ such that one assumes that $Y \sim G(\phi)$ (where $\phi$
represents the distributional parameters), Bayesian inference assumes
that the parameters $\phi$ also follow a probability distribution. The
quantity of interest is the so-called \emph{posterior distribution} of
probable values for $\phi$ given observations $y$ from the random
variable $Y$, which is obtained using Bayes' rule: 
$p( \phi \mid y) \propto p( y \mid \phi) p ( \phi ).$ The uncertainty
of the estimates is then directly obtained from the posterior
distribution $p(\phi \mid y)$. Furthermore, the likelihood $p( y \mid \phi)$
is derived from the model, and has the same mathematical expression as the
likelihood used for MLE methods. Finally, the so-called prior $p(\phi)$ includes
the information known about the parameters $\phi$ \emph{before}
observing the data $y$. For
studies involving extremes, the choice of $p(\phi)$ is of particular
importance, as the small size of the data sample typically results in
a strong influence of the prior over the
posterior. \citep{dey2016_bayes} provides current strategies to choose
appropriate priors when performing inference of the GEV distribution.

For the inference of the parameters in this study, we used a Markov
Chain Monte Carlo (MCMC) sampling scheme. MCMC sampling requires that
the right-hand side of Bayes' rule is known up to a multiplicative
constant, for which it is enough to know the expression for the
likelihood \(p(Y\mid\phi)\) and the prior distribution \(p(\phi)\).

The likelihood term of the DM
approach given by Eqs.~\eqref{eq:glm_mu}-\eqref{eq:glm_xi} is directly
obtained from the GEV distribution (Eq.~\eqref{eq:gev}). For the BR
approach, using the full likelihood is unfeasible as the data's high
dimensionality made the full likelihood intractable; we chose instead
to use the pairwise likelihood from \citep{Padoan2010} (see
Appendix~\ref{appendix:a} for details). The expression for the
pairwise likelihood of the Brown-Resnick model included both the
marginal and dependence parameters so that each MCMC step updated the
value of all
$\phi =\{\rho,\alpha,
\beta_0^{\mu},\beta_0^{\sigma},\beta_0^{\xi},\beta_{i,P}^\psi\}$
parameters simultaneously, where \(\beta_{i,P}^\psi\) denotes all
potential relevant coefficients for $\psi=\left\{\mu,\sigma\right\}$ aside
    from their intercepts $\beta_0^{\mu}$ or $\beta_0^{\sigma}$.

The last step to perform Bayesian inference is to propose a prior
distribution for all parameters. For the DM approach, this includes
the three intercepts $(\beta_0^\mu, \beta_0^\sigma, \beta_0^\xi)$ and
all the possible coefficients $\beta_{i,P}^\psi$, where
$\psi \in \{\mu,\sigma\}$. 

The covariates were recentered around zero so that the value of the
intercepts can be interpreted as the value when all other covariates
are set to their mean values. Based on the study of
\citep{Fischer2017}, who did a similar analysis for the same region,
we use the following priors for the location and scale intercepts:
\begin{align}
\beta_0^\mu &\sim \mathrm{Normal}(1.54,0.6166) \\
\beta_0^\sigma &\sim \mathrm{Normal}(0.4166,0.4166)
\end{align}
The prior for the shape parameter \(\xi\) is a rescaled
Beta-distribution $ \beta_0^{\xi} \sim \mathrm{Beta}(2,2) $ that has
support in \([-0.5,0.5]\). This choice was made as this prior has
already been used by \cite{dyrrdal2015} and also in the operational
application used by MeteoSwiss \citep{meteoswiss}. For the
\(\beta_{i,P}^\psi\), we use the prior
\(\beta_{i,P}^\psi \sim \mathrm{Student-t}(2,0,1)\), which is a
regularizing prior \citep{kruschke2014doing}, preventing overfitting.

For the BR approach, the priors for the marginal response surfaces
were the same as those used for the DM approach
described above. Using the same priors for the two models was done
to simplify the comparison between them. Additionally, the prior for the
range parameter $\rho$ and the smooth parameter $\alpha$
were elicited from typical values of these parameters
in other studies \citep{Zheng2015, Stephenson2016} and were chosen to be
\(p(\rho) = \mathrm{Normal}(30000, 7000)\) and
\(p(\alpha) = \mathrm{Exponential}(2.5)\), respectively. The scales of the
parameters for $p(\rho)$ are in meters. 

MCMC sampling was performed using the software \texttt{Stan} \citep{stan2018}.
A total of 4 chains with 2500 post-warmup samples per
chain using 1000 samples as warmup was used. A visual analysis of the
ridge and trace plots was performed for all models to detect issues with
MCMC chain convergence.

A known issue when using pairwise likelihoods for Bayesian inference
is that the resulting posterior distributions will severely
underestimate the spread of the distribution (\citep{Ribatet2012,
  dey2016_spext, Chan2017}).  The underestimation occurs because the
pairwise likelihood over-uses the data by including each location in
$n/2$ terms of the objective function rather than just one, as would
be the case with the full likelihood, resulting in a likelihood
function that is far too sharply peaked \citep{Ribatet2012}. While
this issue does not severely affect the overall median of the
posterior distribution \citep{Chan2017}, the estimated credible
intervals of the parameters will be strongly underestimated. To tackle
this issue, we applied the \emph{Open Faced-Sandwich} (OFS) correction
proposed by \citep{Shaby2014} to all posterior MCMC samples from the
Brown-Resnick model. The OFS-corrected samples produce credible
intervals that have proper coverage values.  However, it is worth
noting that while the resulting posterior samples fulfill the desired
coverage properties, they are no longer truly
Bayesian. Appendix~\ref{appendix:b} shows a comparison between the raw
MCMC samples and the OFS-corrected ones.

\subsubsection{Prediction of return levels}\label{sec:prediction-of-return-levels}

Once a posterior distribution of the marginal GEV parameters is
obtained from the MCMC samples, it is straightforward to calculate
$q_p(s)$ quantile levels for any probability $p$ of non-exceedance
(i.e., return levels) via the quantile function of the GEV
distribution
\begin{equation}
 q_p(s) = \begin{cases} \mu + \frac{\sigma}{\xi} [(-\log p)^{-\xi} - 1] &
 \xi \neq 0, \\ 
 \mu - \sigma\log(- \log p) &

 \xi = 0\,. \end{cases}
\end{equation}

For each one of the $S$ MCMC sampled parameter values, we calculate a value of $q_p(s)$
with probability $p$. This results in a distribution of $S$ return
levels. We report the median of these return levels as the estimated
return level. Their uncertainty is calculated as the $2.5\%$ and
$97.5\%$ quantiles of the $S$ return levels, forming $95\%$
credibility intervals. Note that the resulting return levels no longer
stem from a GEV distribution but rather from a mixture of many GEV
distributions.

\subsection{Verification and Model Comparison}\label{sec:verification-and-model-comparison}

We use the quantile score (QS) \citep{Bentzien2014} as a measure of
accuracy for both the marginal and the Brown-Resnick models. Given a
series for a single rain gauge of semi-annual block-maxima observations
$(i^l_{d,1}(s_j),...,i^l_{d,N_j}(s_j))$ with $N_j$ years of data for the $j$-th gauge and the corresponding
prediction for the quantile level \(q^l_{p,d}(s_j)\) with probability
$p$ for the same location $s_j$, duration $d$ and season $l$, the QS
is defined as:
\begin{align}
QS^l_{p,d} &= \frac{1}{N} \sum_{t=1}^{N} \rho_p(i^l_{d,t}(s_j) - q^l_{p,d}(s_j)); \\
  &\mathrm{where} \quad \rho_p(u) = [\lvert u \rvert + (2p-1)u]/2.
\label{eq:QS}
\end{align}

The QS is always positive and reaches an optimal value at zero. We
obtain the QS values for both the marginal and the Brown-Resnick model
for probability levels of \(p=(0.9,0.95,0.98,0.99)\), corresponding to
return periods of \((10,20,50,100)\) years.

To compare the performance of two models, \cite{Ulrich2020} defined the
Quantile Skill Index (QSI), a measure derived from the Quantile
Skill Score QSS \citep[cf.][for an introduction to skill scores]{wilks2011statistical}.
Given the QS for a model to be tested ($\mathrm{QS}_\mathrm{model}$) and the QS for a reference
model $(\mathrm{QS}_\mathrm{ref})$, the QSI is defined as
\begin{equation}
  \mathrm{QSI} = \begin{cases} 1-\frac{\mathrm{QS}_{\mathrm{model}}}{\mathrm{QS}_{\mathrm{ref}}}, & \mbox{if }
    \mathrm{QS}_{\mathrm{model}}<\mathrm{QS}_{\mathrm{ref}} \\ -\left(1-\frac{\mathrm{QS}_{\mathrm{ref}}}{\mathrm{QS}_{\mathrm{model}}}\right), &
    \mbox{if } \mathrm{QS}_{\mathrm{model}}\geq
    \mathrm{QS}_{\mathrm{ref}} \end{cases}\, .
\label{eq:QSI}
\end{equation}
Positive (negative) values of the QSI indicate a gain (loss) of skill
for the tested model over the reference. The advantage of the QSI over
the QSS is that the interpretation of negative or positive values is
equivalent (which is not the case for skill scores). For this study,
the tested model is the Brown-Resnick max-stable process model, and
the reference model is the marginal distributional model.

To get an estimation of the out-of-sample performance for the QSI, we
applied 10-fold cross-validation in space to estimate the QS values.
The folds were constructed such that in each one, 90\% of the stations
were used for training the model and the remaining 10\% for
validation. Each station appears in a given validation set once and only
once. This specific cross-validation scheme gives an estimate of how
good the model is at predicting values at ungauged sites, and it does
not give any information on the model's skill at predicting future
observations. Considering the sizeable computational load needed to
perform MCMC sampling for all 8 models for all 10 folds, we opted to
use maximum likelihood instead of Bayesian inference for this
step. Using MLE instead of full Bayesian inference was
considered a fair assessment as we are only interested in point
estimates of return levels when calculating the QS using
Eq.~\eqref{eq:QS}. A separate analysis (not shown) revealed that the
QS point estimates obtained from maximum likelihood were almost always 
very similar to the median QS values obtained from the full posterior
distribution.



\section{Results}

\subsection{Extremal dependence}\label{sec:extremal-dependence}

\begin{figure}
\centering
\includegraphics[scale = 0.8]{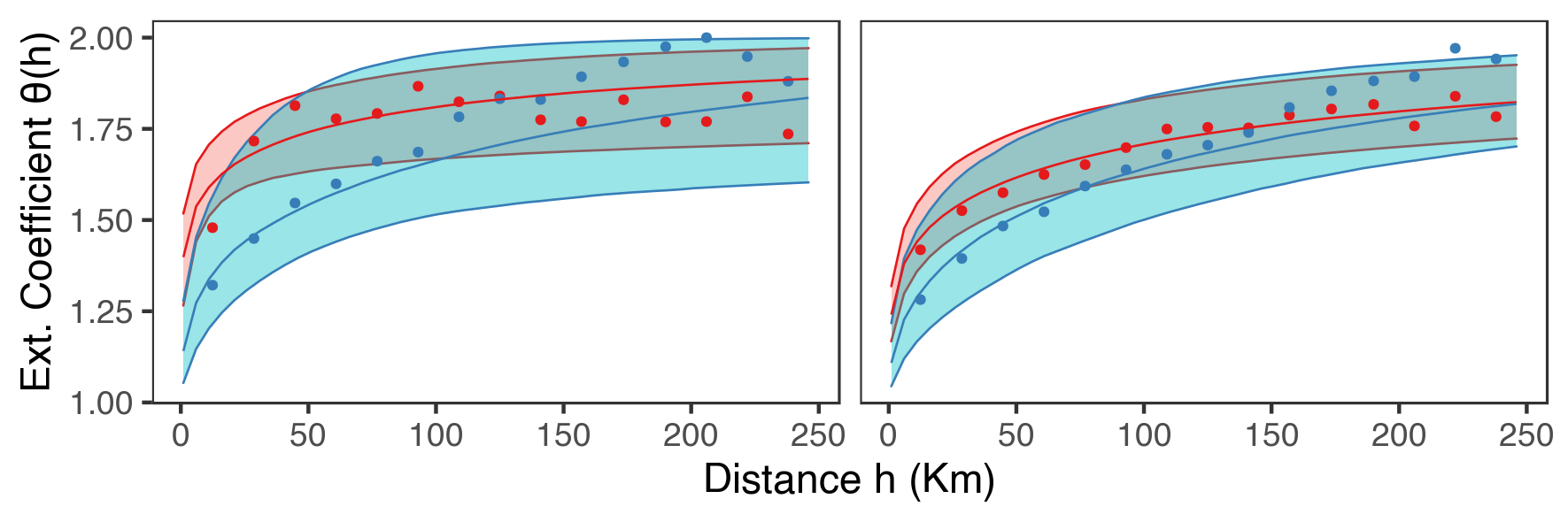}
\caption{Empirical values of the extremal
coefficient $\hat\theta_{\text{NP}}(s_j,s_{j'})$ (dots) and estimated values from the
resulting Brown-Resnick variogram $(\theta_{\mathrm{BR}}(s_j,s_{j'})$ (solid lines,
shaded regions represent the 50\% CI). Colors represent the season: blue
for winter and red for summer. The left panel shows results from the
12-hourly data; the right panel shows results for the daily data.
$\theta(s_j,s_{j'}) \in [1,2]$, where one is complete dependence and two is
complete independence.}
\label{fig:excoef}
\end{figure}

The estimated bivariate extremal coefficient $\hat\theta_{\text{NP}}(s_j,s_{j'})$ for the
$\bm{i}^{\mathrm{(sum,win)}}_{12}(s_j)$ and the $\bm{i}^{\mathrm{(sum,win)}}_{24}(s_j)$ block maxima series
is shown in Fig.~\ref{fig:excoef}. The main feature is that winter maxima (blue) consistently
show lower average values (i.e., higher
extremal dependence) until a distance of around $h = 150$ km. 
For distances $h>150 \mathrm{km}$ this relationship is
inverted. Furthermore, the average distance
where the pairwise maxima still show asymptotical dependence is shown
to be lower than 150 km.  The difference between seasons
is larger for the 12-hour series, possibly reflecting differences in
the rainfall generating mechanisms at this timescale compared to the
24-hour series.

Estimates of the extremal coefficient based on the Brown-Resnick model
($\hat\theta_{\mathrm{BR}}(s_j,s_{j'})$) are also shown in
Fig.~\ref{fig:excoef} as the solid lines with the shaded regions 
representing 50\% credibility
intervals. We compare the values from the Brown-Resnick
model to the empirical $\hat\theta_{\text{NP}}(s_j,s_{j'})$ to get an idea of
how well the BR approach captures the pairwise dependence shown by the data. 
For the 12-hour series (left), this
comparison shows that for winter, the model consistently overestimates
the strength of the dependence for $h>100$, while
for summer, the average dependence is properly captured for
$h \lesssim 150$ km; for greater distances, the dependence is underestimates.
The overestimation in the winter model can also be seen for the daily series (right);
however, it is much less pronounced, with most of the average $\hat\theta_{\text{NP}}(s_j,s_{j'})$
falling inside of the 50\% CIs. 
In both time series, the 50\% CIs are larger for winter; this could suggest that
the extremal dependence for winter is more complex than for summer, resulting in the
winter model exploring a greater range of values for the dependence parameters.
Additionally, the daily series shows less variability than the 12-hour series. This difference
in variability may be due to the increased length of observations for
the daily series. 

An initial inspection of the $\hat\theta_{\text{NP}}(s_j,s_{j'})$ values would suggest that the data shows asymptotic dependence for all series for distances up to $h\leq 150$ km. Therefore, the assumption of asymptotic dependence necessary for using a max-stable process, should be justified. Further discussion about this topic can be found in Appendix \ref{appendix:d}.  

\subsection{Model building}

\subsubsection{Model selection}

The procedure to choose the orders for the Legendre Polynomials of
Eqs.~\eqref{eq:glm_mu}-\eqref{eq:glm_xi} results in the models
described in Tab.~\ref{tab:orders}. Basic prior and posterior
predictive checks were performed to detect any misspecification
issues; some examples for the reference stations
described below can be found in Appendix~\ref{appendix:c}.
A visual analysis of the prior and posterior checks did not detect any severe
issues.  
\begin{table}[htbp]
\centering
\caption{Maximum chosen orders of Legendre Polynomials for the distributional model
  in Eqs.~\eqref{eq:glm_mu}-\eqref{eq:glm_sigma}}.
\begin{tabular}{lll}
\hline
             & $\mu$ & $\sigma$ \\ \hline
summer (24h) & 2     & 1        \\
winter (24h) & 3     & 2         \\
summer (12h) & 2     & 1          \\
winter (12h) & 3     & 2       \\ \hline
\end{tabular}
\label{tab:orders}
\end{table}

\subsection{Parameter estimates}

Parameter estimates for the dependence and shape parameters are
reported in Tab.~\ref{tab:estparameters} as 
median and 95\% credibility intervals for each parameter. A
distinct difference can be seen in the value of the range parameter
$\rho$ (in meters) between summer and winter, as the value in winter is always
significantly larger than for summer, regardless of the time
scale. This result is consistent with the behavior of the extremal
coefficient seen in Fig. \ref{fig:excoef}, and it may indicate that
the rainfall events leading to the block maxima in winter are, on
average, larger than those in summer. Furthermore, the shape parameter
shows a difference for winter and summer, regardless of the time scale.

\begin{table}[h]
  \centering
  \caption{Bayesian estimates of the Brown-Resnick max-stable model
    parameters. Posterior medians are reported along with their 95\%
    credible interval limits on either side as (lower,median,upper). The
      coefficients corresponding to the Legendre Polynomials were
      omitted from this table.}
  \label{tab:estparameters}
  \begin{tabular}{llll}
    \hline
    & $\rho$            & $\alpha$        & $\xi$              \\ \hline
    12h (s)& 413,4896,11997    & 0.17,0.40,0.64  & 0.05,0.18,0.31  \\
    12h (w)& 3596,43870,104192 & 0.31,0.78,1.24  & -0.01,0.08,0.20 \\
    24h (s)& 4722,22993,43936  & 0.39,0.54,0.69  & 0.15,0.24,0.33  \\
    24h (w)& 6801,53228,109265 & 0.57,0.83,1.12  & -0.01,0.09,0.21 \\ \hline
  \end{tabular}
\end{table}

\subsection{Marginal parameters and return levels}

\begin{figure}
\centering
\includegraphics[width=\textwidth]{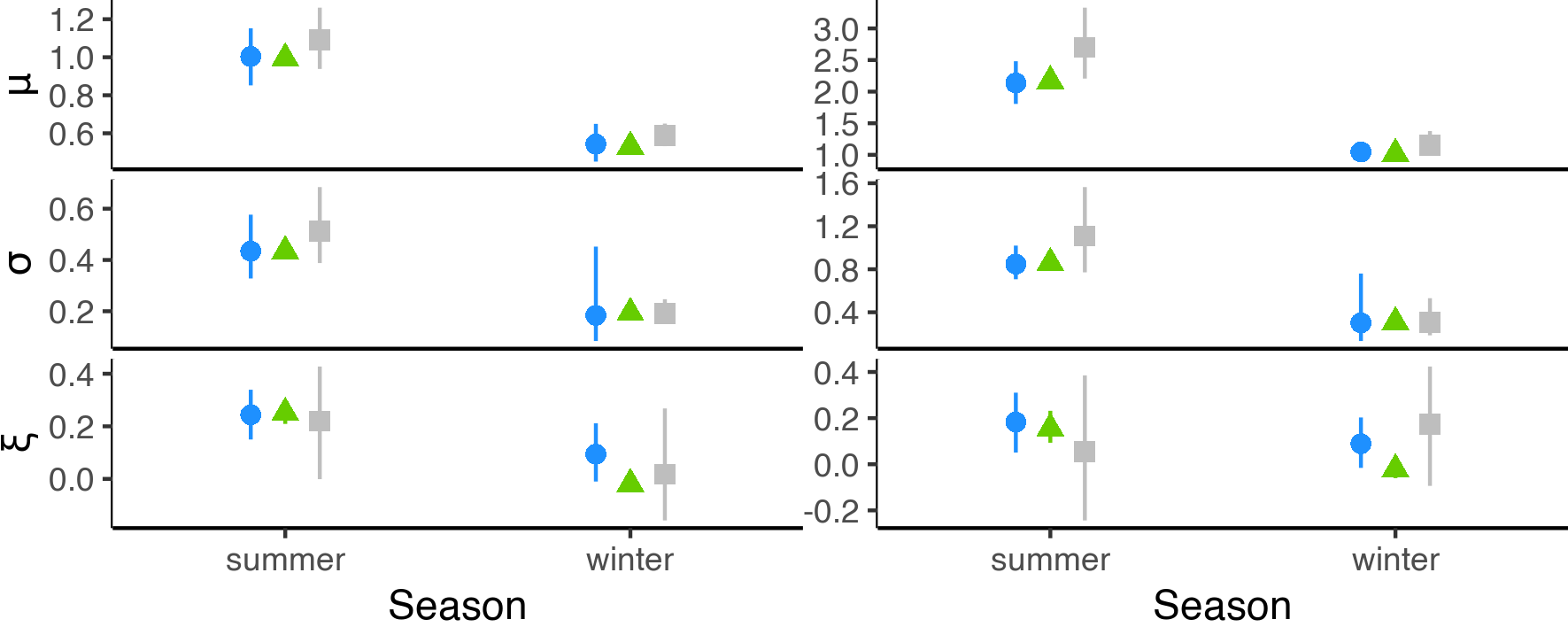}
\caption{Estimated values of the
location \(\mu\), scale \(\sigma\), and shape \(\xi\) parameters of the
GEV distribution for station Potsdam. The symbols' shape indicates the model
used for estimation: circle (blue) = BR, triangle (green) = DM, square
(gray) = pointwise GEV.}
\label{fig:potsdam_parameters}
\end{figure}

\begin{figure}
\centering
\includegraphics[width=\textwidth]{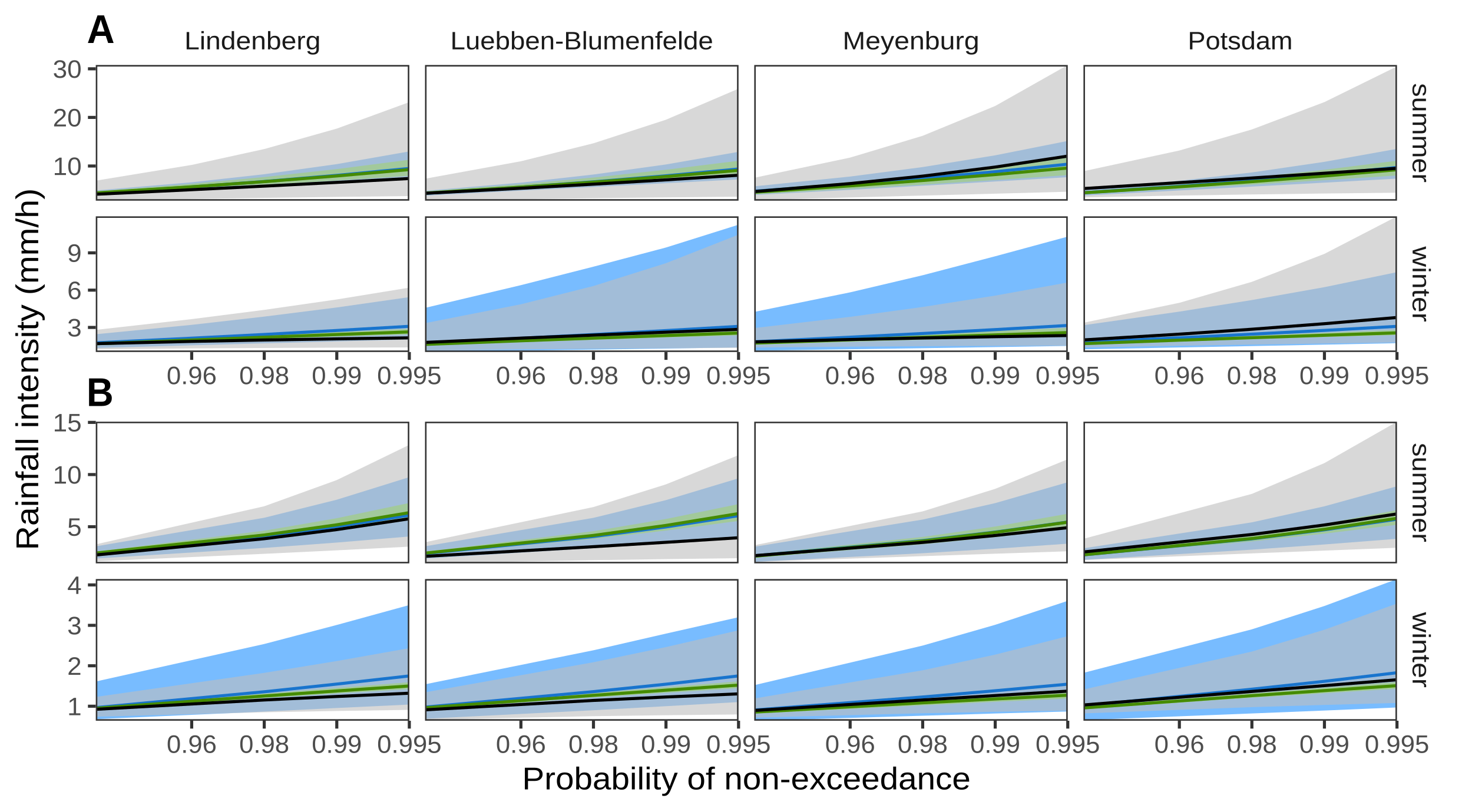}
\caption{Return level of precipitation intensity (mm/h). Color denotes 
the model used: Blue for the BR model, green for the DM model, and gray 
for the pointwise GEV. Shaded regions represent pointwise 95\% 
credibility intervals. (A) 12-hour data, (B) daily data. For reference,
the probabilities of non-exceedance $p = (0.96,0.98,0.99,0.995)$ correspond to the
$(25,50,100,200)$ year return periods, respectively. }
\label{fig:return-levels}
\end{figure}

Four reference stations were chosen to illustrate the differences
  in the marginal GEV parameters and return levels
from the DM and the BR models (respective locations of the reference
stations are given by red
diamonds in Fig.~\ref{fig:BB_map}). We chose the two stations with the
longest time series, which are closely surrounded by other stations
(Potsdam and Lindenberg), a station with a long time series that is
isolated from other stations (Meyenburg), and a station with a short
time series which is surrounded by other stations
(Luebben-Blumensfelde). Figures~\ref{fig:potsdam_parameters}-\ref{fig:return-levels}
show the GEV parameter estimates and the resulting return levels with
95\% credibility intervals, respectively. Furthermore, pointwise GEV
estimates and their resulting return levels with 95\% credibility
intervals were added for reference; these estimates were obtained
using the same priors for the intercepts described in section
\ref{sec:inference-for-both-models}.

Concerning the GEV parameters, Fig.~\ref{fig:potsdam_parameters} shows
that the pointwise estimates (taken here as the median value of the
posterior distributions) are similar for the DM and the BR
models. This similarity was expected, as the marginal parameters are
only vaguely affected by the spatial dependence through their
incorporation in the likelihood term of Eq.~\eqref{eq:pairwise}.
However, when comparing the models, a pattern concerning the
uncertainty of the estimated parameters (taken here to be the 95\%
credibility intervals) is visible.  For summer (and in the case of
$\xi$, also winter), the highest uncertainty is always seen for the
pointwise GEV model, followed by the BR model and the DM model, which
consistently show the smallest uncertainty. In contrast, the largest
uncertainty for location and scale in winter can be seen for the BR
model, followed by the pointwise GEV and the distributional
models. This phenomenon can be observed in other stations (not
shown). We infer that the uncertainty estimated for the marginal
parameters is strongly affected by the underlying spatial dependence,
which changes according to the rainfall-generating mechanisms dominant
in the respective season.

To further delve into the last point, Fig.~\ref{fig:return-levels}
presents how the return levels for different non-exceedance
probabilities for the BR and DM approaches differ. As before, we
compare different seasons and two different durations. The median
return level is generally similar across the different models, with
increasing differences for larger probabilities of non-exceedance.  In
contrast, the uncertainty is noticeably different for each model,
which is consistent with the results of the GEV parameters.  In
summer, the uncertainty is always largest for the
pointwise GEV model, followed in order by the BR and the DM
models. This changes in winter, when the uncertainty is largest
typically for the BR model, with a few exceptions.  Surprisingly, it
would appear that the inclusion of the max-stable dependence on the
model for winter resulted in an overall increase in the uncertainty,
even when compared to the pointwise model that contains no information
about other stations.  This result may be associated with a loss of
skill for the BR model when modeling block maxima in winter, an aspect
that will be explored in the next section.

\subsection{Model comparison}

We now explore how the seasonal differences in the extremal dependence
affect the accuracy of the return levels estimates using the BR model.
We use the DM approach as reference in the QSI to assess how
much the dependence influences the return level estimates.  Positive
(negative) QSI values mean that the predicted return levels for
ungauged sites have better (worse) QS values for the dependent BR
model than for the independent DM one. For this study, our main focus
is on the QSI difference between seasons, as we believe this arises
from a change in the extremal dependence when analyzing the
semi-annual block maxima from different meteorological regimes.

\begin{figure}
\centering
\includegraphics[width=\textwidth]{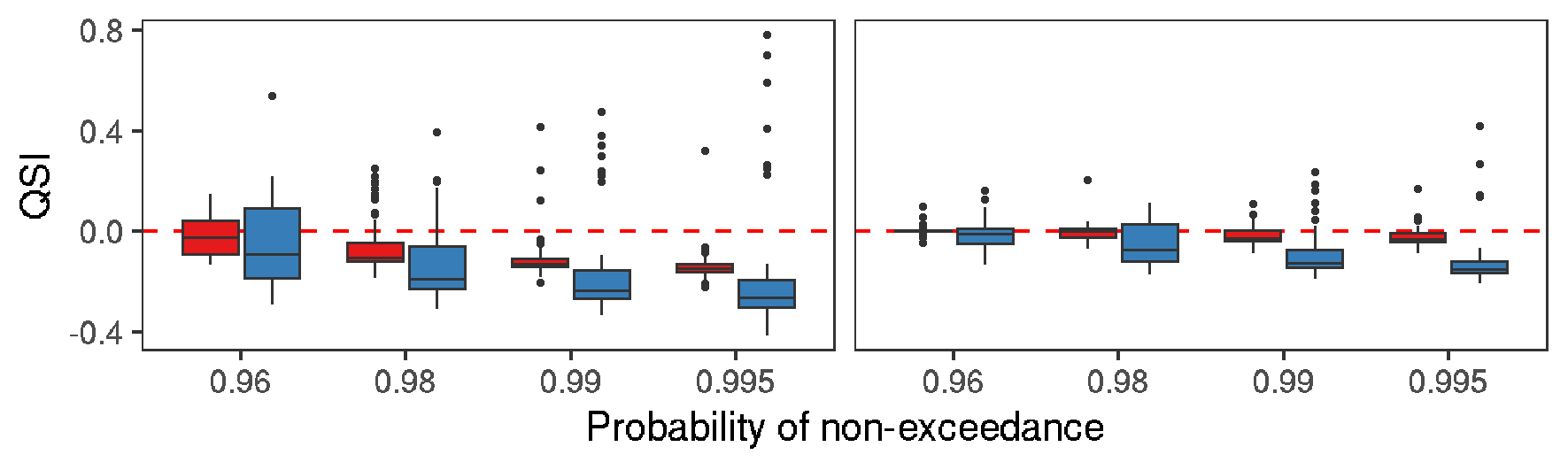}
\caption{Boxplots showing the distribution of the Quantile Skill Index
  for all stations for 12-hourly (left) and daily (right) data. The
  colors indicate the season. Positive (negative) values indicate an
  increase (decrease) in skill for the BR approach compared to the DM
  one.}
\label{fig:QSIboth}
\end{figure}

Figure~\ref{fig:QSIboth} depicts the distribution of the cross-validated
QSI values over all stations. The
12-hourly data shows an overall average loss of skill for the
winter and summer models when using the BR model.  This loss of skill
increases as the non-exceedance probability increases, with the winter
model showing substantially lower average QSI values than the summer
model.  Furthermore, the variability in QSI values is noticeably
larger for winter than summer; in fact, the highest QSI value is
always found within an outlier of the winter model. For the daily
data, the winter model shows the same decrease in skill with
increasing non-exceedance probabilities with high variability;
however, in this case, the summer model consistently shows QSI values
close to zero with very low variability.  Finally, a noteworthy
difference between average QSI values can be observed between summer
and winter for both periods. This difference increases with the
non-exceedance probability, but it remains constant between the
12-hourly and daily periods.

\begin{figure}
\centering
\includegraphics[width=0.9\textwidth]{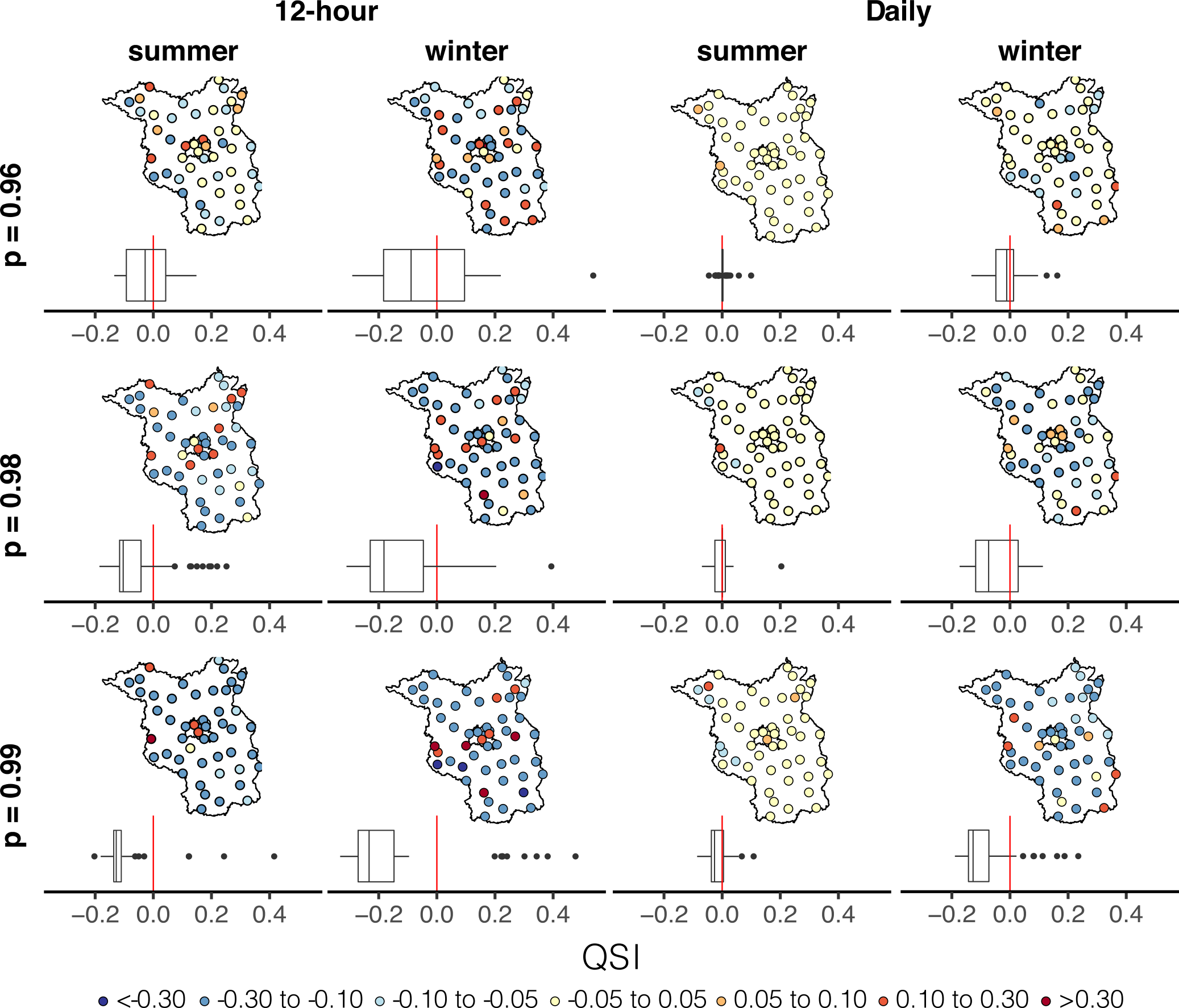}
\caption{Spatial distribution within Berlin-Brandenburg (solid line)
  of QSI values for the 12-hourly (leftmost two columns) and daily
  (rightmost two columns) data. Boxplots below each map show the
  distribution of QSI values as in Fig.~\ref{fig:QSIboth}}
\label{fig:QSImap}
\end{figure}

To further explore the difference in QSI values for both seasons and
durations, Figure~\ref{fig:QSImap} shows the spatial distribution of
QSI values, i.e.\ values for every station. No apparent pattern is
visible from the different configurations, as QSI values appear to be
largely random. Nevertheless, the lowest QSI appear mostly at stations close to
the domain's border, suggesting that the BR model performs better when
a station is surrounded on all sides by other stations. This effect is
admittedly not very reliable, as stations with very low values of QSI
can also be found within the middle of the domain. A closer inspection
of the difference between seasons reveals a subtle pattern: similar
QSI values seem to cluster in summer, while the distribution is
predominantly random in winter. This change could be attributed to the
difference in the rainfall generating processes, as will be discussed
in the following.


\section{Discussion}

The results described in the last section provide compelling evidence
that the extremal dependence shown by the data changes sufficiently
enough to have a noticeable effect in the resulting marginal estimates
down the line when using a model capable of capturing such dependence
(the BR model in our case). This difference was mainly observed when
comparing different marginal quantities from two seasons: the
estimated GEV parameters (with their respective return levels) and the
cross-validated Quantile Score (QS), an out-of-sample performance
measure for the predicted return levels for ungauged sites.

The observed difference in marginal estimates when using a spatial
model is consistent with previous extreme rainfall studies;
\cite{Stephenson2016}, for example, reports that the incorporation of
the max-stable process dependence led to an overall shift towards
heavier tailed marginal distributions across their study
location. They also found that the uncertainty for the estimated
marginal quantities was larger for the max-stable process model than
for the independent model, which is in good agreement with our results
for summer maxima estimated return levels. Additionally, the spatial
distribution of the QSI falls in line with \cite{Le2018}, who found
that return levels estimated from a max-stable process presented
noticeable differences in their spatial distributions compared to an
unconditional model. However, these studies did not estimate the
impact of this difference on the model performance.  Our study then
provides insight into the operational use of Brown-Resnick max-stable
models by first examining how different types of rainfall-generating
mechanisms affect the marginal estimates and then applying a model
validation framework for the
out-of-sample ungauged model accuracy.

A comparison of the uncertainty in the GEV parameters and the return
levels showed that when modeling summer maxima, the uncertainty
resulting from the BR model appeared to be a middle point between the
DM and the pointwise GEV models. However, the significant reduction in
uncertainty for the DM model compared to the pointwise GEV model
signals that this model underestimates the natural variability in the
rainfall data.  Thus, it seems plausible that the larger uncertainty
seen in the BR model is a more accurate representation of such
variability.  In contrast, when dealing with winter maxima, the
uncertainty obtained by the BR model appears to have been consistently
overestimated compared to the pointwise approach. From our results, it
is not completely clear why this is the case, but it may be speculated
that the isotropic Brown-Resnick dependence model was misspecified for
the extremal dependence structure in the winter data. A possible
source of error is the assumption that the dependence structure is
isotropic, which might be a better approximation for convective
events than for synoptic/mixed events that occur in
winter. On the other hand, the larger values estimated
for the range and smooth parameters indicate that the dependence is
stronger for winter than in summer; however, these parameters do not
say anything about the isotropic/anisotropic structure. This larger
dependence in winter could be attributed to frontal events being
generally larger and more elongated than convective events. 
Thus, more stations are simultaneously affected
by the same event, increasing the
dependence. Figure~\ref{fig:numEvents} in appendix~\ref{appendix:d}
reports how many unique events resulted in block maxima being chosen
from the daily series in winter and summer. This table supports the
idea that the events are larger in winter, as the number of unique
events is consistently lower in winter than in summer. However,
Fig.~\ref{fig:excoef} also reveals a surprising increase in dependence
for distances larger than 120 km; this may suggest that some
underlying weather patterns from a larger scale than the convective
scale influence the dependence.

Our findings report that the Brown-Resnick model is mostly as good as
the unconditional DM model when modeling summer block maxima, whereas
the BR model presents a remarkable loss in skill compared to the DM
model when modeling winter block maxima. It is worth noting that past
studies have primarily focused on summer maxima, as the convective
nature of the rainfall-generating mechanisms in this season typically
leads to the annual maxima events to occur in summer. Our findings
suggest that the isotropic Brown-Resnick dependence model is a proper
first approximation when dealing with block maxima resulting from convective events. On the
other hand, the loss in skill for the winter maxima model provides
further evidence that this model is misspecified when dealing with
either synoptic, stratiform, or a mixture of synoptic/convective
events.

We acknowledge potential limitations to this study.  An important
question for future studies is to determine the effect of anisotropy
in the results, which, as discussed above, is expected to have an
important role in modeling the spatial dependence for synoptic
events. Furthermore, previous studies have shown that rain gauge
networks are typically too scarce to resolve convective cells properly
\citep{Lengfeld2019}. Thus, in order to get a better representation of
the spatial dependence, future work should make use of radar networks
to complement rain gauge data.  A significant limitation of our work
was the use of the pairwise likelihood instead of the full likelihood
of the Brown-Resnick model within the Bayesian framework. While some
of the most known issues with this approach were tackled by using the
Open-Faced Sandwich approach of \citep{Shaby2014}, it would be
beneficial instead to use a full-likelihood approach such as that of
\citep{Dombry2017}. Furthermore, due to the high computational demand
of performing Cross-Validation within a Bayesian setting, the QS and
QSI results reported in the results come from a maximum likelihood
estimation.  Moreover, we assumed that the data was stationary,
ignoring the possible effects of climate change. The effect of this
non-stationarity on the extremal dependence should be explored in
further studies, as it has been shown that accounting for
non-stationarity results in a measurable effect on the return level
estimates \citep{Ganguli2017}.  Our study indirectly classified
precipitation types based on dominant types for different
seasons. Further studies should use a direct classification of event
types, which would avoid the mixing of convective and frontal events
in winter. Some work in classifying extreme events already exists, for
example, that of \citep{lengfeld2021}. Furthermore, the use of
max-stable processes requires that the data present asymptotic tail
dependence, an assumption that does not hold for aggregation durations
lower than 12 hours. For a more in-depth study of convective events
shorter durations would be needed; in this case, a more flexible model
that can capture both asymptotic tail dependence and independence
would be needed, such as the one proposed by
\citep{wadsworth2019higher}, which was applied to hourly rainfall data
by \citep{jordan2022}.

This study indicates that different rainfall mechanisms can
strongly influence the spatial dependence presented by the block
maxima. This change in the dependence structure can, in turn, result
in significant misspecification of the model if not accounted for
properly. Thus, it is essential to understand the types of
rainfall-generating mechanisms in the domain of study when using
max-stable models.

\section*{Acknowledgments}

OEJ and HR acknowledge support from the Deutsche
Forschungsgemeinschaft (DFG) within the research training program
NatRiskChange (GRK 2043/1). HR and MO acknowledge support from the
German Federal Ministry of Education and Research (BMBF) through the
ClimXtreme project, grant numbers 01LP1902H and 01LP1902I, respectively. OEJ acknowledges support
from the Mexican National Council for Science and Technology (CONACyT)
and the German Academic Exchange Service (DAAD). Finally, we thank
Jana Ulrich for her input on the statistical model design. Our
acknowledgment does not imply endorsement of our views by these
colleagues, and we remain solely responsible for the view expressed
herein.

\section*{Code Availability}
All data and code is available in the following Code Ocean repository: \url{https://doi.org/10.24433/CO.9114783.v2}

\section*{Authors' contributions}

Conceptualization: OEJ, MO, HR; Methodology: OEJ, MO, HR; Formal analysis and investigation: OEJ; Software - OEJ; Visualization: OEJ; Writing - original draft preparation: OEJ; Writing - review and editing: MO, HR; Funding acquisition: HR; Resources: HR; Supervision: MO, HR.


\appendix

\section{Inference from the Brown-Resnick max-stable process}\label{appendix:a}

Inference is done using the pairwise log-likelihood \citep{Padoan2010}, which for our study is
\begin{equation}
L(\phi \mid i^l_{d,1}(s_1), ..., i^l_{d,N}(s_J)) = \sum_{t = 1}^{N} \sum_{j = 1}^{J-1} \sum_{j'=j+1}^{J} \log f(i^l_{d,t}(s_j),i^l_{d,t}(s_{j'})\mid\phi),
\label{eq:pairwise}
\end{equation}
where $\phi =\{\rho,\alpha, \beta_0^{\mu},\beta_0^{\sigma},\beta_0^{\xi},\beta_{i,P}^\psi\}$  
represents the parameters to estimate, $i^l_{d,t}(s_j)$ is the observed semi-annual block maxima for the duration $d$ and season $l$ at location $s_j$ for year $t$, and each term $f(\cdot,\cdot)$ is 
the appropriately transformed bivariate density function derived from the bivariate
distribution function for the Brown-Resnick process given by
\begin{multline}
\mathrm{Pr}[Z(s_1) \leq z_1, Z(s_2) \leq z_2] = \\ \exp \left[
  -\frac{1}{z_1} \Phi \left( \frac{\sqrt{\gamma(h)}}{2} +
    \frac{1}{\sqrt{\gamma(h)}} \log \frac{z_2}{z_1} \right) - 
  \frac{1}{z_2} \Phi \left( \frac{\sqrt{\gamma(h)}}{2} +
    \frac{1}{\sqrt{\gamma(h)}} \log \frac{z_1}{z_2} \right) \right]\, .
\label{eq: BR_pair_cdf}
\end{multline}
Here $z$ follows a unit Fréchet distribution, $\Phi$
denotes the standard normal distribution function, $h$ is defined as
the euclidean distance between $s_1$ and $s_2$, and the
variogram $\gamma$ is defined in Eq. ~\ref{eq:variogram}. In equation~\eqref{eq:pairwise} it is assumed that the number of years $N$ is equal for all $J$-stations. However, this is not the case, as some stations have longer records than others. We took $N$ to be the one from the station with the longest records, and whenever a station did not have data for the $t$-th year, we made the corresponding term in the log-likelihood to be zero. However, the time period used for all stations was chosen to minimize the number of paired stations with no data.

\section{Number of events per year}\label{appendix:d}

\begin{figure}[h!]
\centering
\includegraphics[width=0.9\textwidth]{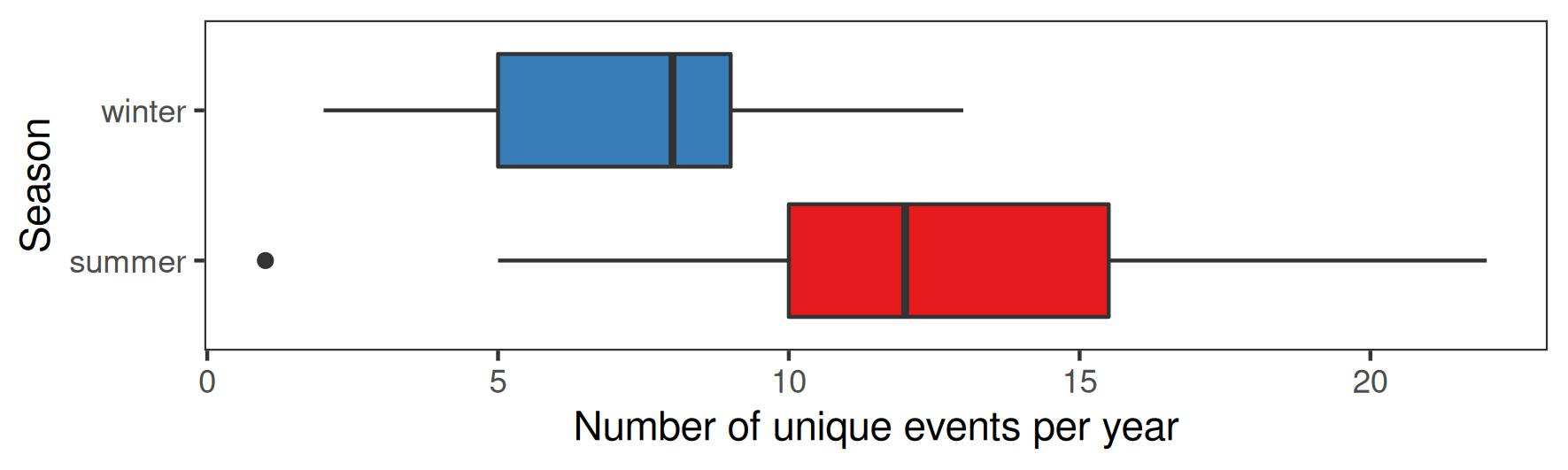}
\caption{Boxplots showing the difference in unique events for the different seasons studied. Only the 24 hour data is shown.}
\label{fig:numEvents}
\end{figure}

Figure \ref{fig:numEvents} shows how many unique events that resulted in block maxima 
were seen from the daily $i^{\mathrm{(sum,win)}}_{24,l}$ series. Overall, the number
of unique events in summer is larger than in winter.

\section{OFS correction for Bayesian Inference using composite likelihood}\label{appendix:b}

\begin{figure}[h!]
\centering
\includegraphics[width=0.9\textwidth]{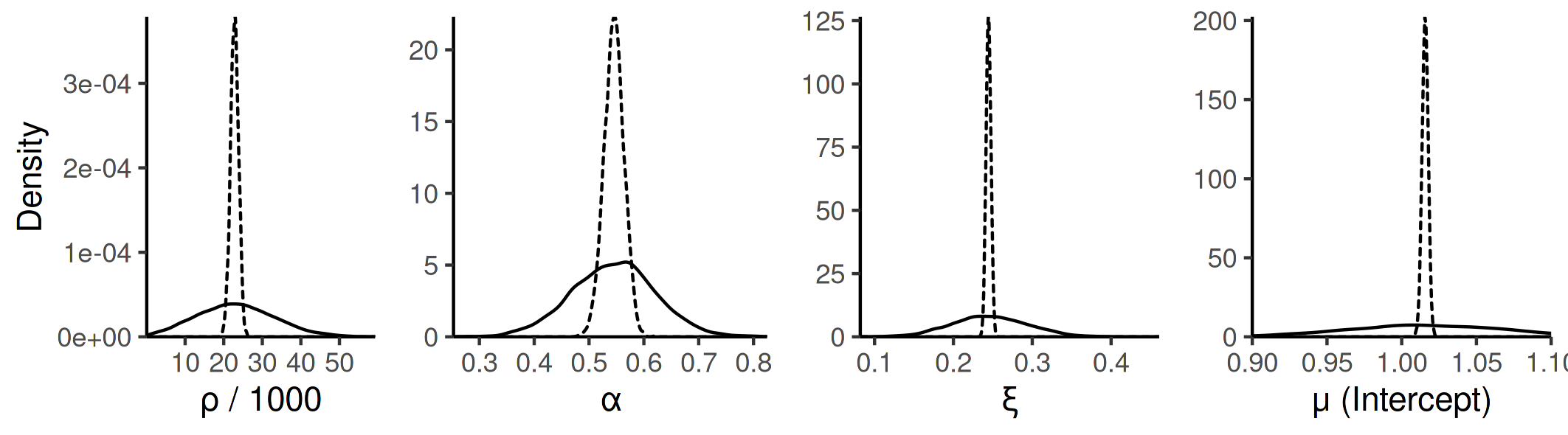}
\caption{Density plots for the raw MCMC samples (dashed line) and the resulting OFS
corrected samples (continuous lines) for 4 selected parameters from the daily summer
results. }
\label{fig:OFS}
\end{figure}

A comparison between the uncorrected raw samples from the MCMC sampling
using Stan with the pairwise likelihood of Eq. \eqref{eq:pairwise} and the
corresponding samples corrected using the Open-Faced Sandwich (OFS)
correction from \citep{Shaby2014} is shown in Fig. \ref{fig:OFS}. It can be seen
that the uncorrected samples grossly underestimate the uncertainty shown
by the 95\% credibility intervals. On the other hand, the OFS corrected samples keep the
same median but ``stretch'' the resulting uncertainty so
that the desired 95\% coverage of the intervals is achieved.

\section{Analysis of asymptotic dependence using extremal coefficient}\label{appendix:d}

\begin{figure}[h!]
\centering
\includegraphics[width=0.9\textwidth]{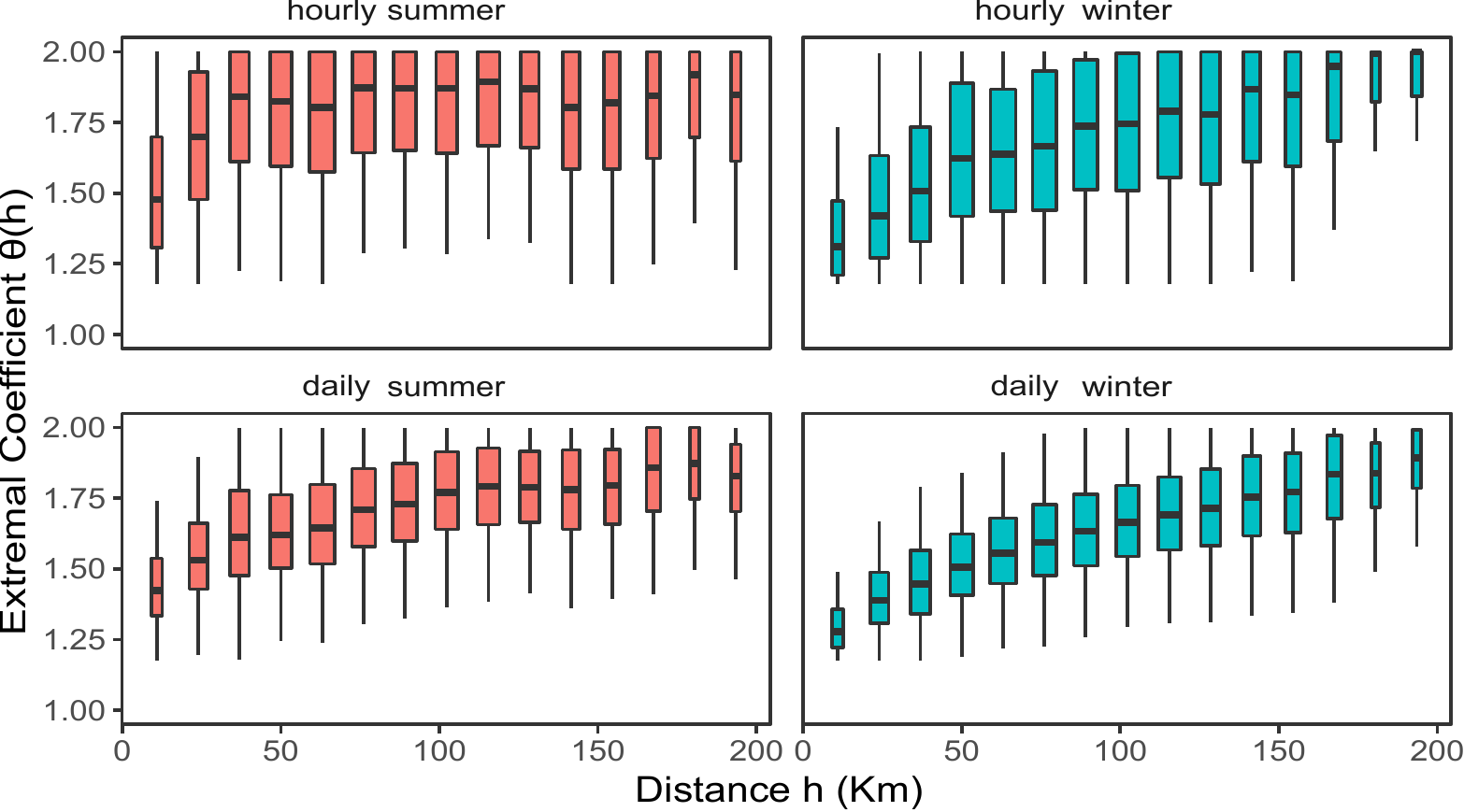}
\caption{Boxplots showing the distribution of bootstraped samples (N=500) of the non-parametrical estimate of the extremal coefficient $\hat{\theta}_\mathrm{NP}(s_j,s_{j'})$. The width of the boxplots is proportional to the amount of data.}
\label{fig:App_D}
\end{figure}

Figure \ref{fig:App_D} shows the distribution of bootstraped samples for $\hat{\theta}_\mathrm{NP}(s_j,s_{j'})$, where the estimation method is the same as the one used for Fig.~\ref{fig:excoef}. The bootstraped samples provide an estimate of the uncertainty that allows us to judge the asymptotic dependence conditions present in the data. 

The figure shows that for the daily series, both seasons show a value of the extremal coefficient below 1.75 for $h \leq 150$ km, suggesting that the data is asymptotically dependent at least for this distance. After 150 km, the coefficient goes close to 2, but not immediately. On the other hand, the situation is different between summer and winter for the hourly frequency. Here, the uncertainty is much larger, which could be a reflection of the smaller number of years. Furthermore, while the winter series behaves similar to the daily winter series having reasonably strong dependence for distances up to 150 km, the hourly summer series tends very quickly to lower dependence levels. This again suggests that the events in summer are typically smaller in size that those in winter. Asymptotical dependence can be reasonably suggested for the hourly winter data, but for hourly summer, one could argue this is true only for relatively short distances. However, the uncertainty is rather large, wit a lot of values still falling under the strong dependence case. Therefore, we make the assumption for asymptotical dependence for all four series. 

\section{Model diagnostic and posterior predictive checks for reference stations}\label{appendix:c}

\begin{figure}[h!]
\centering
\includegraphics[width=0.9\textwidth]{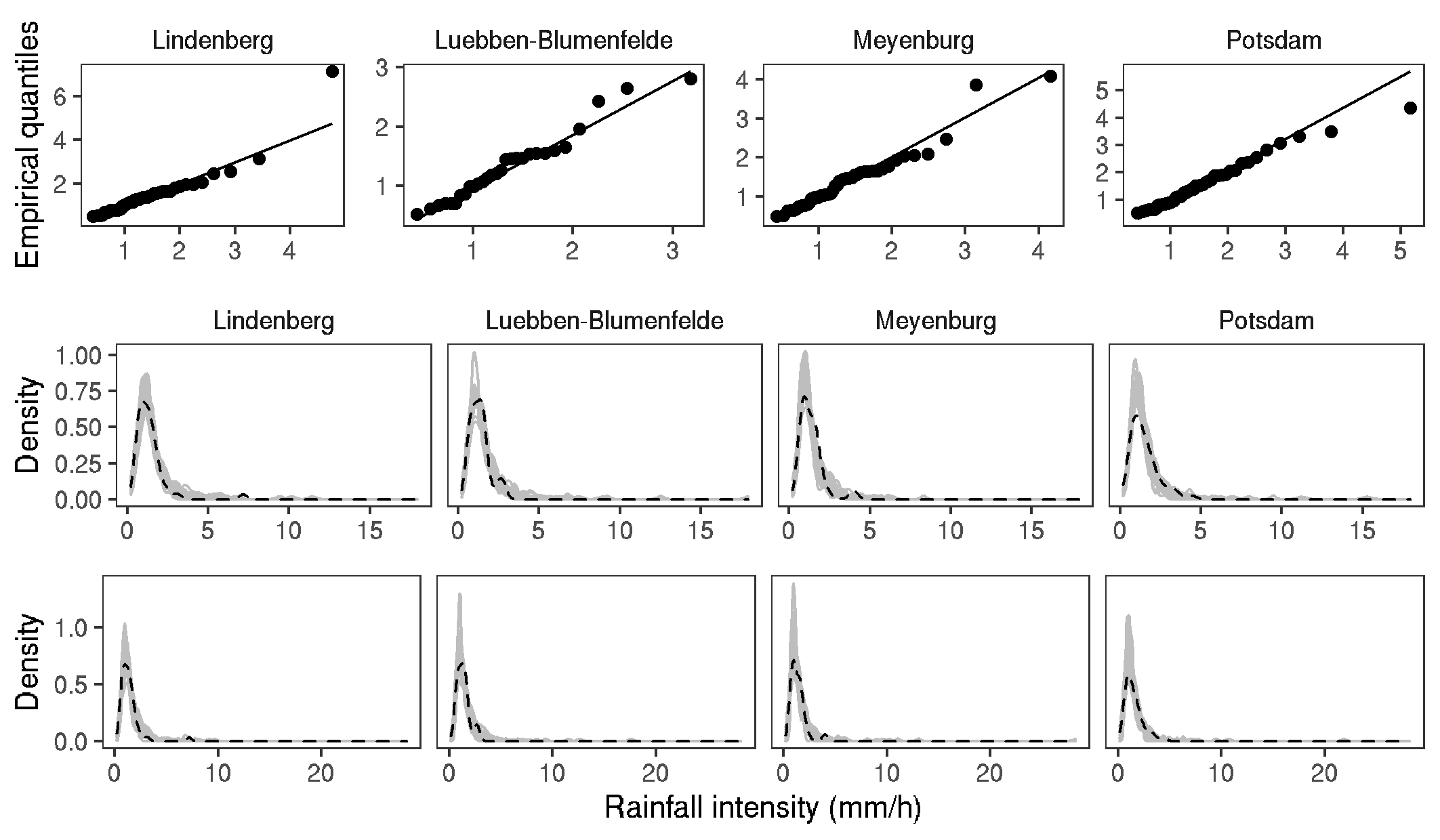}
\caption{Top row: Q-Q plots for the four reference stations. Middle row: Posterior
predictive check for the DM model. Bottom row: Posterior predictive check for the BR model. For the middle and bottom row, the dashed line shows the observed density, and the grey lines show 100 samples from 20 GEV distributions with parameters sampled from the respective posterior distributions.}
\label{fig:App_C}
\end{figure}

We obtained Quantile-Quantile plots and Posterior predictive checks to assure that our model adequately represents the observed data. Some of these are shown in Fig. \ref{fig:App_C}. The included plots come from the hourly summer data. 

The QQ plots for the different stations provide evidence that the GEV is mostly appropriate for modeling the marginal distributions, with some minor exceptions. Similarly, the posterior predictive checks allow us to see how well the posterior distributions for the DM and BR models would be at capturing the original data. In this case, both models' original data seems well-captured.

\bibliographystyle{unsrtnat}
\bibliography{references}

\end{document}